\documentclass[10pt]{iopart}
\usepackage[latin1]{inputenc}
\usepackage{amssymb}
\usepackage{amsfonts}
\usepackage{latexsym}
\usepackage{theorem}
\usepackage{bbm}
\usepackage[nosort]{cite}

\newtheorem{define}{Definition}[section]
\newtheorem{theo}[define]{Theorem}
\newtheorem{lemma}[define]{Lemma}
\newtheorem{propo}[define]{Proposition}
\newtheorem{coro}[define]{Corollary}

\newcommand{\ket}[1]{\ensuremath{|#1 \rangle}}
\newcommand{\bra}[1]{\ensuremath{\langle #1|}}

\newcommand{\kb}[1]{\ensuremath{| #1 \rangle \! \langle #1 |}}

\newcommand{\ip}[2]{\ensuremath{\langle #1 | #2 \rangle}}
\newcommand{\op}[2]{\ensuremath{| #1 \rangle \! \langle #2 |}}

\newcommand{\1}{\ensuremath{\mathbbm{1}}}
\newcommand{\F}{\ensuremath{\mathbb{F}}}

\newcommand{\abs}[1]{\ensuremath{| #1 |}}
\newcommand{\cb}[1]{\ensuremath{\| #1 \|_{cb}}}
\newcommand{\norm}[1]{\ensuremath{\| #1 \|_{\infty}}}
\newcommand{\tracenorm}[1]{\ensuremath{\| #1 \|_{1}}}
\newcommand{\hsnorm}[1]{\ensuremath{\| #1 \|_{2}}}

\newcommand{\C}{\ensuremath{\mathbb{C}}}
\newcommand{\R}{\ensuremath{\mathbb{R}}}
\newcommand{\N}{\ensuremath{\mathbb{N}}}

\newcommand{\supp}[1]{\ensuremath{{\rm supp }(#1)}}
\newcommand{\rank}[1]{\ensuremath{{\rm rank }(#1)}}
\newcommand{\trace}[1]{\ensuremath{{\rm tr}(#1)}}
\newcommand{\ld}{\ensuremath{{\rm ld} \, }}
\newcommand{\id}{\ensuremath{{\rm id} \, }}

\newcommand{\re}[1]{\ensuremath{{\rm Re}(#1)}}
\newcommand{\im}[1]{\ensuremath{{\rm Im}(#1)}}

\newcommand{\hh}{\ensuremath{\mathcal{H}}}
\newcommand{\bh}{\ensuremath{\mathcal{B(H)}}}
\newcommand{\bhh}[1]{\ensuremath{\mathcal{B}(\mathcal{H}_{#1})}}

\newcommand{\bhstar}{\ensuremath{\mathcal{B_{*}(H)}}}
\newcommand{\bhhstar}[1]{\ensuremath{\mathcal{B}_{*}(\mathcal{H}_{#1})}}

\newcommand{\kk}{\ensuremath{\mathcal{K}}}

\newcommand{\bkkstar}{\ensuremath{\mathcal{B_{*}(K)}}}

\newcommand{\bkkkstar}[1]{\ensuremath{\mathcal{B}_{*}(\mathcal{K}_{#1})}}

\begin{document}

\title{{\em Tema con variazioni:} quantum channel capacity}

\author{Dennis Kretschmann and Reinhard F Werner}

\address{Institut f\"ur Mathematische Physik, Technische
Universit\"at Braunschweig, Mendelssohnstr. 3, 38106 Braunschweig,
Germany}

\ead{d.kretschmann@tu-bs.de}

\begin{abstract}
Channel capacity describes the size of the nearly ideal channels,
which can be obtained from many uses of a given channel, using an
optimal error correcting code. In this paper we collect and
compare minor and major variations in the mathematically precise
statements of this idea which have been put forward in the
literature. We show that all the variations considered lead to
equivalent capacity definitions. In particular, it makes no
difference whether one requires mean or maximal errors to go to
zero, and it makes no difference whether errors are required to
vanish for any sequence of block sizes compatible with the rate,
or only for one infinite sequence.
\end{abstract}


\submitto{\NJP}

\section{Introduction, Outline and Notations}
\label{sec:introduction}

Quantum channel capacity is one of the key quantitative notions of
the young field of quantum information theory. Whenever one asks
``how much quantum information'' can be stored in a device, or
sent on a transmission line, it is implicitly a question about the
capacity of a channel. Like Shannon's classical definition, the
concept applies also to noisy channels, which do corrupt the
signal. In this case one may apply an error correction scheme, and
still use the channel like an almost ideal one. Capacity expresses
this quantitatively: it is the maximal number of ideal qubit
(resp. bit) transmissions per use of the channel, taken in the
limit of long messages and using error correction schemes
asymptotically eliminating all errors.

Many of the terms in this informal definition can be, and in fact
have been, formalized mathematically in different ways. As a
result, there are many published definitions of quantum capacity
in the literature. Some of these are immediately seen to be
equivalent, but with other variants this is less obvious.
Moreover, some of the differences seem to have gone unnoticed,
creating the danger that some results would be unwittingly
transferred between inequivalent concepts, creating a mixture of
rigorous argument and folklore hard to unravel.

The purpose of the present paper is to show that, fortunately, all
the major definitions are indeed equivalent. In order to make the
presentation self-contained, we have also included abridged
versions of arguments from the literature. Other points, however,
e.g., concerning the question whether the rate has to be achieved
on every sequence of increasing blocks, or just on an infinite,
possible sparse set of increasing block sizes, seem to be new. We
have also made an effort to lay out the required tools carefully,
so that they can be used in other applications.

All this does not help much to come closer to the proof of a
coding theorem, i.e., to a rigorous formula for the capacity not
requiring the solution of asymptotically large optimization
problems. Major progress in this direction has recently been
obtained by Shor \cite{Sho02,Sho03} and Devetak \cite{Dev03}. We
hope that our work will contribute to an unambiguous
interpretation of these results, as well.\\
\\
The key chapter (Section~\ref{sec:tema}) of this paper begins with
presenting the theme: a basic rigorous definition of quantum
channel capacity. This is followed by nine logical variations on
this theme, which like musical variations are not all of the same
weight. In each variation a result is stated to the effect that a
modified definition is equivalent to the basic one after all. All
proofs, however, are left to the later sections. A Coda at the end
of the variations comments on the coding theorem and recent
developments.

\begin{itemize}
 \item[2.\hskip8pt]Tema: Quantum Channel Capacity
 \item[2.1.]Choice of units
 \item[2.2.]Testing only one sequence
 \item[2.3.]Minimum fidelity
 \item[2.4.]Average fidelity
 \item[2.5.]Entanglement fidelity
 \item[2.6.]Entropy rate
 \item[2.7.]Errors vanishing quickly or not at all
 \item[2.8.]Isometric encodings and homomorphic decodings
 \item[2.9.]Little help from a classical friend
 \item[2.10.]Coda: The Coding Theorem
 \end{itemize}

 \vskip12pt
\noindent{\bf Notations}\\ In order to state the basic definition
of capacity and its variations, we have to introduce some
notation. A {\em quantum channel} which transforms input systems
described by a Hilbert space $\hh_1$ into output systems described
by a (possibly different) Hilbert space $\hh_2$ is represented by
a completely positive trace-preserving linear map $T \mathpunct :
\; \bhhstar{1} \rightarrow \bhhstar{2}$, where by $\bhstar$ we
denote the space of trace class operators on $\hh$. This map takes
the input state to the output state, i.\,e., we work in the
Schr\"odinger picture (see Kraus' textbook \cite{Kra83} for a
detailed description of the concept of quantum operations).

The definition of channel capacity requires the comparison of the
channel after correction with an ideal channel. As a measure of
the distance between two channels we take the {\em norm of
complete boundedness} (or {\em cb-norm}, for short) \cite{Pau02},
denoted by $\cb{\cdot}$. For two channels $T,S$, the distance
$\frac12\cb{T-S}$ can be defined as the largest difference between
the overall probabilities in two statistical quantum experiments
differing only by exchanging one use of $S$ by one use of $T$.
These experiments may involve entangling the systems on which the
channels act with arbitrary further systems. Equivalently, we may
set $\cb T=\sup_n\norm{T\otimes\id_n}$, where the norm is the norm
of linear operators between the Banach spaces $\bhhstar{i}$, and
$\id_n$ denotes the identity map (ideal channel) on the $n\times
n$ matrices.

Among the properties which make the  cb-norm well-suited for
capacity estimates are multiplicativity,  $\cb{T_1 \otimes T_2} \,
= \, \cb{T_1} \, \cb{T_2}$, and unitality, $\cb{T} \, = \, 1$ for
any channel $T$. The equivalence with other error measures is
discussed extensively below.

Note that throughout this work we use base two logarithms, and we
write $\ld x := \log_2 x$.


\section{{\em Tema:} Quantum Channel Capacity}
\label{sec:tema}

\begin{define}
\label{define:basic}
    A positive number $R$ is called {\bf achievable rate} for the
    quantum channel $T \mathpunct : \bhhstar{1} \to \bhhstar{2}$
    with respect to the quantum channel $S \mathpunct : \; \bhhstar{3}
    \to \bhhstar{4}$ iff for any pair of integer sequences $(n_\nu)_{\nu
    \in\N}$ and $(m_\nu)_{\nu \in \N}$ with $\lim_{\nu\to\infty} n_\nu \,
    = \, \infty$ and $\overline{\lim}_{\nu\to\infty}
    \frac{m_\nu}{n_\nu} \, \leq \, R$ we have
    \begin{equation}
       \lim_{\nu\to\infty} \, \Delta(n_\nu,m_\nu) \, = \, 0,
    \end{equation}
    where we set
    \begin{equation}
        \Delta(n_\nu,m_\nu) \, := \, \inf_{D,E} \, \cb{D \, T^{\otimes n_\nu}
        E \, - \, S^{\otimes m_\nu} \,},
    \end{equation}
    the infimum taken over all encoding channels $E$ and decoding channels
    $D$ with suitable domain and range.\\
    The {\bf channel capacity} $Q(T,S)$ of $T$
    with respect to $S$ is defined to be the supremum of all
    achievable rates. The {\bf quantum capacity} is the special case
    $Q(T) := Q(T,\id_2)$, with $\id_2$ being the ideal qubit channel.
\end{define}
This definition is a transcription of Claude E. Shannon's
definition of the capacity of a discrete memoryless channel in
classical information theory, as presented originally in his
famous 1948 paper \cite{Sha48} and now found in most standard
textbooks on the subject (e.g., \cite{Mce77,Ash90}). To make the
translation one only needs to express Shannon's maximal error
probabilities in terms of norm estimates \cite{Wer01} and take an
ideal one-bit channel rather than the one-qubit channel as the
reference. This choice can also be made for quantum channels $T$,
defining the capacity $C(T)$ of a quantum channel for {\it
classical information}. Much more is known about $C(T)$ than about
$Q(T)$ \cite{BS98,Sho00}.


\subsection{{\em Prima Variazione:}  Choice of Units}
\label{sec:ideal1}

Formally, Definition \ref{define:basic} assigns a special role to
the ideal qubit channel $\id_2$. Is this essential? What do we get
if we take the ideal channel $\id_n$ on a Hilbert space of some
dimension $n>2$ as reference?

We will show in Section \ref{sec:ideal2} that the choice $n = 2$
only amounts to a choice of units, fixing the unit {\em bit}:
\begin{equation}
    Q(T, \id_n) \, = \, \frac{\ld m}{\ld n} \, Q(T, \id_m).
\end{equation}


\subsection{{\em Seconda Variazione:} Testing Only One Sequence}
    \label{sec:sequences1}

At first sight Definition \ref{define:basic} of channel capacity,
as given above and widely used throughout the community
\cite{Wer01,HW01,Ham01,Key02,ON02}, seems a little impractical,
since it involves checking an infinite number of pairs of
sequences when testing a given rate $R$. Work would be
substantially reduced if only one such pair had to be tested. For
the sake of discussion let us say that a rate $R$ is {\it
sporadically achievable} if, for {\it some} pair of sequences
$(n_\nu)_{\nu \in \N}$, $(m_\nu)_{\nu \in \N}$, with
$n_\nu\to\infty$ and vanishing errors, the rate $R$ is achieved
infinitely often: $\overline{\lim}_{\nu\to\infty} \,
\frac{m_\nu}{n_\nu} \, = \, R$. For example, there might be a
special coding scheme which utilizes some rare number theoretical
properties of $n$ and $m$. Many published definitions
\cite{BNS98,BKN00,BST98,Ham02} would accept sporadically
achievable rates as achievable in the capacity definition. Often
the choice $n_\nu = \nu$ is made
\cite{BDS97,DSS98,HHH00,MU02,KW02}. While this sequence of block
sizes can hardly be called ``sporadic'', it is a logically similar
variation to the sparse sequences, so we include it for
convenience.

In Section~\ref{sec:sequences2} we show that all sporadically
achievable rates are, in fact, achievable. Hence there is no need
to introduce a ``sporadic capacity''. What we have to show is that
coding schemes that work infinitely often can be extended to all
block sizes. This is a non-trivial result, since we also show that
by merely putting blocks together, and by perhaps not using some
of the code bits such an extension is not possible.


\subsection{{\em Terza Variazione:} Minimum Fidelity}
\label{sec:subspace}

The cb-norm is by no means the only way to evaluate the distance
between two channels. Another distance measure that has appeared
particularly widely (e.\,g., in \cite{BDS+96,BDS97,DSS98,BKN00})
is the minimal overlap between input and corresponding output
states: The {\em minumum fidelity} of a quantum channel $T
\mathpunct : \; \bhstar \, \rightarrow \, \bhstar$ is defined as
\begin{equation}
\label{equation:subspace}
    F(T) \, := \, \min \left \{ \, \bra{\psi} \, T (\kb{\psi}) \, \ket{\psi} \; \mid \; \psi \in \hh, \|
    \psi \| = 1 \, \right \}.
\end{equation}
When we want to particularly emphasize the Hilbert space $\hh$ on
which the minimization is performed, we will write $F(\hh, T)$
instead.

Of course, $0 \leq F(\hh,T) \leq 1$, and $F(\hh,T) = 1$ implies
that $T$ acts as the ideal channel on $\hh$: $T \hspace{-0.2cm}
\mid_{\hh} \; = \id_{\hh}$. These features make the minimum
fidelity a suitable distance measure. We might then call a
positive number $R$ an achievable rate for the channel $T$ if
there is a sequence $(\kk_n)_{n \in \N}$ of Hilbert spaces such
that $\overline{\lim}_{n\to\infty} \; \frac{\ld \dim (\kk_n)}{n} =
R$ and $\lim_{n\to\infty} \, F(\kk_n, D_n \, T^{\otimes n} \, E_n)
= 1$ for suitable encodings $(E_n)_{n \in \N}$ and decodings
$(D_n)_{n \in \N}$.

In Section \ref{sec:equidist} we show that the quantum channel
capacity arising from this definition is the same.


\subsection{{\em Quarta Variazione:} Average Fidelity}
\label{sec:average1}

Instead of requiring that the maximum error be small we might be
less demanding, and just require an average error to vanish. In
the previous section we would then have to replace the minimum
fidelity $F(T)$ by the {\em average fidelity},
\begin{equation}
\label{equation:average}
    \overline{F}(T) \, := \, \int \bra{\psi} \, T
    (\kb{\psi}) \, \ket{\psi} \; d\psi,
\end{equation}
where the integral is over the normalized unitarily invariant
measure ``$d\psi$'' on the unit vectors in $\hh$.

In Sections \ref{sec:averagefidel} and \ref{sec:entmin} we show
that this modification has no effect on the quantum channel
capacity. An alternative proof is presented in Section
\ref{sec:average2}.


\subsection{{\em  Quinta Variazione:} Entanglement Fidelity}
    \label{sec:entanglement}

{\em Entanglement fidelity} was introduced by Ben Schumacher in
1996 \cite{Sch96} and is closely related to minimum fidelity. It
characterizes how well the entanglement between the input states
and a reference system not undergoing the noise process is
preserved: For a quantum channel $T \mathpunct : \; \bhstar \,
\rightarrow \, \bhstar$ and a quantum state $\varrho \, \in \,
\bhstar$, the {\em entanglement fidelity of $\varrho$ with respect
to $T$}, $F_e(\varrho,T)$, is given as
\begin{equation}
\label{equation:entanglement}
    F_e(\varrho,T) \, := \, \bra{\psi} \, T \otimes \id_\bhstar \,
    (\kb{\psi})\, \ket{\psi},
\end{equation}
where $\psi$ is a purification of $\varrho$. This quantity does
not depend on the details of the purification process, as is made
evident by the alternative expression \cite{Sch96}
\begin{equation}
\label{equation:entanglement24x}
    F_e(\varrho,T) \, = \, \sum_i | {\rm tr} \varrho \, t_i |^2,
\end{equation}
where $T(\sigma)=\sum_i t_{i}^{} \sigma t_i^*$ is the Kraus
decomposition of $T$ \cite{Kra83}. Obviously, $0 \, \leq \,
F_e(\varrho,T) \, \leq \, 1$. Moreover, $F_e(\varrho,T) \, =  \,
1$ implies that $T$ is noiseless on the support of $\varrho$:
 $T\!\mid_{\supp \varrho} \, = \, \id_{\supp \varrho}$.

We might then define achievable rates exactly as in Section
\ref{sec:subspace} above, replacing the condition
$\lim_{n\to\infty} \, F(\kk_n, D_n \, T^{\otimes n} \, E_n) = 1$
by the requirement that
\begin{equation}
    \label{equation:entanglement24y}
        \lim_{n\to\infty} \, \inf_{\varrho\in\bkkkstar{n}} \,
        F_e(\varrho, D_n \, T^{\otimes n} \, E_n) = 1
\end{equation}
for suitable encodings $(E_n)_{n \in \N}$ and decodings $(D_n)_{n
\in \N}$.

The quantum capacity that stems from this definition of achievable
rates is likewise equivalent, as shown in Section
\ref{sec:equidist}.\\
\\
In the previous definition, instead of minimizing over all density
operators $\varrho \in \bhstar$, one can simply choose $\varrho$
to be the maximally mixed state on $\hh$, $\varrho := \frac{1}{d}
\1_{\hh}$, with the shorthand $d := \dim \hh$. The resulting
variant of entanglement fidelity we call {\em channel fidelity}
\cite{RW03} of the quantum channel $T$,
\begin{equation}
    \label{equation:chanfid}
        F_{c}(T) \, := \, F_e(\frac{1}{d} \1_{\hh},T) \,
        = \, \bra{\Omega} (T \otimes \id_{\hh}) \, (\kb{\Omega}) \, \ket{\Omega},
\end{equation}
where $\Omega = d^{-1/2} \, \sum_{i=1}^{d} \ket{i,i}$ is a
maximally entangled state on $\hh \otimes \hh$.

Channel fidelity is a very handy figure of merit, since it is a
linear functional, does not involve a maximization process, and is
completely equivalent to the error criteria discussed above. The
details are spelled out in Section~\ref{sec:averagefidel}.\\
\\
A further variant arises when in the definition of channel
fidelity instead of the maximally entangled state $\Omega$ an
arbitrary input state $\Gamma \in \hh \otimes \hh$ is permitted,
replacing the channel fidelity $F_{c}(T)$ by the quantity
$F_{c}(\Gamma, T) := \bra{\Omega} (T \otimes \id_{\hh}) \,
(\kb{\Gamma}) \, \ket{\Omega}$. This is the error quantity I.
Devetak's {\em entanglement generating capacity} \cite{Dev03} is
built on, and is likewise equivalent (Section \ref{sec:entgen}).


\subsection{{\em  Sesta Variazione:} Entropy Rate}
\label{sec:entrate}

The original definition of quantum channel capacity in terms of
entanglement fidelity involves a different concept of computing
the rates \cite{BNS98,BST98}. According to this definition, the
capacity of a quantum channel is the maximal {\em entropy rate} of
a {\em quantum source} whose entanglement with the reference
system is preserved by the noisy channel. A quantum source
$(\kk_n,\varrho_n)_{n \in\N}$ consists of a pair of sequences of
Hilbert spaces $\kk_n$ and corresponding density operators
$\varrho_n \, \in \, \bkkkstar{n}$. It is meant to represent a
stream of quantum particles produced by some physical process. Its
{\em entropy rate} is defined as
\begin{equation}
    \label{equation:entropyrate}
        R=\overline{\lim}_{n\to\infty} \, \frac1n{S(\varrho_n)}\, ,
\end{equation}
where $S(\varrho) \, = \, - \, \trace{\varrho \, \ld \varrho}$ is
the {\em von Neumann entropy}.

The quantum capacity  as defined by Schumacher is then the
supremum of all entropy rates for sources such that
$\lim_{n\to\infty} \, F_e(\varrho_n, D_n \, T^{\otimes n} \, E_n)
= 1$ for suitable encodings $(E_n)_{n \in \N}$ and decodings
$(D_n)_{n \in \N}$.

It turns out that in order to make this definition equivalent to
the others, some mild constraint on the sources is needed. In
fact, we will show in Section \ref{sec:entmin} that the supremum
over {\it all} sources will be infinite for all channels with
positive capacity. However, for a wide range of interesting
sources equivalence does hold, namely (cf. Section
\ref{sec:entmin}),
\begin{itemize}
\item if $\rho_n=\1_{\kk_n}/\dim\kk_n$, which brings us back to
the definition based on channel fidelity discussed in the previous
section,

\item if the source satisfies the so-called asymptotic
equipartition property, which has recently been established for
general stationary ergodic quantum sources
\cite{BKS+02,BKS+03,BS03}, or

\item if the dimension of the ambient space of the encodings grows
at most exponentially (even at a rate much larger than the
capacity).
\end{itemize}


\subsection{{\em Settima Variazione:} Errors vanishing quickly or not at all}
\label{sec:speed}

In the various definitions of achievable rates presented so far,
instead of simply requiring the error quantity to approach zero in
the large block limit, one could impose a certain minimum speed of
convergence, e.\,g., linear, polynomial, exponential, or
super-exponential convergence, as a function of the number of
channel invocations. We will show in Section \ref{sec:sequences2}
that all these definitions coincide, as long as the  speed of
convergence is at most exponential. \\
\\
If we require the errors to vanish even faster or, in the extreme
case, that $\Delta(n_\nu,m_\nu)\, = \, 0$ for large enough $\nu$,
as in theory of error correcting codes invented by Knill and
Laflamme \cite{KL97}, equivalence no longer holds: If a channel
has a small, but non-vanishing probability for depolarization, the
same also holds for its tensor powers, and no such channel allows
the perfect transmission of even one qubit. Hence the capacity
based on exactly vanishing errors will be zero for such
channels.\\
\\
On the other hand, one might sometimes feel inclined to tolerate
(small) finite errors in transmission: For some $\varepsilon > 0$,
let $Q_{\varepsilon}(T)$ denote the quantity defined exactly like
the quantum channel capacity in Definition \ref{define:basic}, but
requiring only $\Delta(n_\nu,m_\nu) \leq \varepsilon$ for large
$\nu$ instead of $\lim_{\nu\to\infty} \, \Delta(n_\nu,m_\nu) \, =
\, 0$. Obviously, $Q_{\varepsilon}(T) \, \geq \, Q(T)$ for any
quantum channel $T$. We even have $\lim_{\varepsilon\to 0}
Q_{\varepsilon}(T) \, = \, Q(T)$ (see Section
\ref{sec:sequences2c}).

In the purely classical setting even more is known: If
$\varepsilon \, > \, 0$ is small enough, one cannot achieve bigger
rates by allowing small errors, i.\,e., $C_{\varepsilon}(T) \, =
\, C(T)$. This is the so-called {\em strong converse} to Shannon's
coding theorem. It is still unknown whether an analogous property
holds for quantum channels.


\subsection{{\em Ottava Variazione:} Isometric Encodings and Homomorphic Decodings}
\label{sec:restrict}

Definition \ref{define:basic} of channel capacity involves an
optimization over the set of all encoding and decoding maps. This
set is very large, and it may thus seem favorable to restrict both
encoding and decoding to smaller classes.

In \cite{BKN00} it has been shown that we may restrain our
attention to {\em isometric} encodings, i.\,e., encodings of the
form
\begin{equation}
\label{equation:isometric}
    E(\varrho) \, = \, V \, \varrho \, V^*
\end{equation}
with isometric $V$, and still be left with the same capacity (see
Section \ref{sec:isometric} for details). Physically, this means
that encoding can always be thought of as a unitary process
augmented by an initial projection onto a subspace small enough to
fit into the channel.\\
\\
In the Knill/\,Laflamme setting of perfect error correction
\cite{KL97}, not only are encoding maps isometric, but in addition
the decodings can be chosen to be of the (Heisenberg picture) form
\begin{equation}
\label{equation:homomorphic}
    D_{*}(X) \, = \, V \, (X \otimes \id) \, V^* \, + \,
    \trace{\varrho_0 \, X} \, (\id - V \, V^*)
\end{equation}
with isometric $V$ and an arbitrary reference state $\varrho_0$.
We call maps of this type {\em homomorphic}, because the first
term is an algebraic homomorphism, and the second term only serves
to render the whole channel unital.

Since the sufficiency of isometric encoding transfers from the
perfect error correction setting to asymptotically perfect error
correction, it may seem reasonable to conjecture that a similar
result holds for homomorphic decodings. However, up to now no such
result is known.


\subsection{{\em Nona Variazione:} Coding with a Little Help from a
Classical Friend}
    \label{sec:friend}

Here we consider a setup in which a quantum channel $T$ is
assisted by additional classical forward communication between the
sender (Alice) and the receiver (Bob). Clearly, this allows Alice
and Bob to collaborate in a more coordinated fashion: Alice may
use the additional resource to transfer information about the
encoding process, which Bob on his part may try to take advantage
of in his choice of the decoding channel.

However, it is a straightforward consequence of the isometric
encoding theorem that these new possibilities do not help to
increase the channel capacity, even if the classical side channel
is noiseless: We have $Q(T \otimes \id_c) = Q(T)$
\cite{BDS+96,BKN00}, where by $\id_c$ we denote an ideal channel
of arbitrarily large dimensionality. That this is not a trivial
statement is seen from the observation that classical feedback
between successive channel uses may increase the capacity
\cite{BDS+96,BDS97,BST98,Bow02}.

The uselessness of classical forward communication may be extended
to cover so-called {\it separable} side channels, i.\,e., quantum
channels with intermediate measurement and re-preparation
processes. The details on both classes of side channels are
spelled out in Section \ref{sec:sideinfo}.


\subsection{{\em Coda:} The Coding Theorem}
\label{sec:coda}

Computing channel capacities on the basis of the definitions
given, even the simplified ones, is a tricky business. It involves
optimization in systems of asymptotically many tensor factors. It
has therefore been a long-time challenge to find a quantum
analogue of Shannon's noisy coding theorem \cite{Sha48}, which
would allow to compute the channel capacity as an optimization
over a low dimensional space.

According to Shannon's famous theorem, the classical capacity is
obtained by finding the supremum of the so-called {\em mutual
information}, which itself is given in terms of the Shannon
entropy. A quantum analogue of mutual information, {\em coherent
information}, has been identified early. For the quantum channel
$T \mathpunct : \, \bhhstar{1} \to \bhhstar{2}$ and the density
operator $\varrho \in \bhhstar{1}$, it is defined as
\begin{equation}\label{equation:CohInf}
   I_c (\varrho, T) \, := \, S(T(\varrho)) \, - \, S \left (T \otimes \id
   \, (\kb{\psi}) \right ),
\end{equation}
where $\psi \in \hh_1 \otimes \hh_1$ is a purification of the
density operator $\varrho$, and $S$, as before, is the {\em von
Neumann entropy}.

The regularized coherent information has long been known to be an
upper bound on the quantum channel capacity
\cite{BNS98,BST98,BKN00}, i.e.,
\begin{equation}
    \label{equation:cohinf2}
        Q(T) \leq \lim_{n\to\infty} \frac{1}{n} \max_{\varrho} I_c
        (\varrho, T^{\otimes n}).
\end{equation}
Unlike the classical or quantum mutual information, coherent
information is not additive; hence taking the limit $n\to\infty$
in Equation (\ref{equation:cohinf2}) is indeed required
\cite{DSS98}.

The first sketch of an argument how to close the gap in Equation
(\ref{equation:cohinf2}) was given by Seth Lloyd \cite{Llo97}. At
a recent conference, Peter Shor presented \cite{Sho02,Sho03} a
coding scheme based on random coding to attain coherent
information. His results have not been published yet. Shortly
thereafter, Igor Devetak released \cite{Dev03} another coding
scheme based on a key generation protocol made ``coherent''
\cite{Har03,DHW03}. By the same techniques Devetak and Winter
\cite{DW03a,DW03b} very recently were able to prove the
long-conjectured {\em hashing inequality} \cite{BDS+96}, which
states that the regularized coherent information is an achievable
entanglement distillation rate, and implies the channel capacity
result by teleportation \cite{HHH00}.

These achievements certainly mark a major step in the direction of
a coding theorem, but do not satisfy all the properties desired of
such a theorem. In particular, they still demand the solution of
asymptotically large variational problems.


\section{Elementary Properties of Channel Capacity}
\label{sec:properties}
\subsection{Basic Inequalities}
\label{elementary}

Before we enter the proof sections we need to review some basic
properties of channel capacity, which will turn out to be helpful as we
proceed, but are also interesting in their own right. All proofs are
easy, and may be found in \cite{Key02}, albeit for noiseless reference
channels only. The generalization is straightforward.

Running two channels, $T_1$ and $T_2$, in succession, the capacity
of the composite channel, $T_1 \circ T_2$, cannot be bigger than
the capacity of the channel with the smallest bandwidth. This is
known as the {\em bottleneck inequality}:
\begin{equation}
\label{equation:bottleneck}
    Q(T_1 \circ T_2, S) \, \leq \, \min \, \{\, Q(T_1,S), Q(T_2,S) \}.
\end{equation}
Instead of running $T_1$ and $T_2$ in succession, we may also run them in
parallel, which is represented mathematically by the tensor product $T_1
\otimes T_2$. In this case the capacity can be shown to be {\em
super-additive},
\begin{equation}
\label{equation:superadditive}
    Q(T_1 \otimes T_2, S) \, \geq \, Q(T_1,S) \, + \, Q(T_2,S).
\end{equation}
For the standard ideal channels we even have additivity. The same
holds true if both $S$ and one of the channels $T_1, T_2$ are
noiseless, the third channel being arbitrary. However, to decide
whether additivity holds generally is one of the big open problems
in the field.\\
\\
Finally, the {\em two step coding inequality} tells us that by
using an intermediate channel in the coding process we cannot
increase the transmission rate:
\begin{equation}
\label{equation:twostep} Q(T_1,T_3) \, \geq \, Q(T_1,T_2) \;
Q(T_2,T_3).
\end{equation}


\subsection{Quantum Capacity of Noiseless Channels}
    \label{sec:ideal2}

There are special cases in which the quantum channel capacity can
be evaluated relatively easily, the most relevant one being the
noiseless channel $\id_n$, where by the subscript $n$ we denote
the dimension of the underlying Hilbert space. In this case we
have
\begin{equation}
    \label{equation:ideal}
        Q(\id_n,\id_m) \, = \, \frac{\ld n}{\ld m}.
\end{equation}
A proof follows below. Combining this with the two-step coding
inequality (\ref{equation:twostep}), we see that for any quantum
channel $T$
\begin{equation}
    \label{equation:units}
        Q(T,\id_n) \, = \, \frac{\ld m}{\ld n} \, Q(T,\id_m),
\end{equation}
which shows that quantum channel capacities relative to noiseless
quantum channels of different dimensionality only differ by a
constant factor. Fixing the dimensionality of the reference
channel then only corresponds to a choice of units. Conventionally
the ideal qubit channel is chosen as a standard of reference,
fixing the unit {\em bit}.\\

 \noindent{\bf Proof of Equation (\ref{equation:ideal})} This is an immediate
consequence of estimates of the simulation error
$\Delta(\id_n,\id_m)=\inf_{D,E}\cb{D \, \id_n E-\id_m}$ between
ideal channels. We have
\begin{eqnarray}
   \Delta(\id_n,\id_m) &=  0\;,
               \qquad\ \ & \mbox{ if }\quad m \leq n;
\label{delta0}   \\
   \Delta(\id_n,\id_m) &\geq 1 - \frac{n}{m}\;,\
                 & \mbox{ if }\quad m \geq n.
\label{equation:noiseless4}
\end{eqnarray}
The first relation is shown by explicitly constructing
 $E \mathpunct : \;\mathcal{B}_{*}(\C^m) \rightarrow \mathcal{B}_{*}(\C^n)$
and
 $D \mathpunct : \; \mathcal{B}_{*}(\C^n) \rightarrow\mathcal{B}_{*}(\C^m)$
such that $D\, \id_n \, E=\id_m$. To this end we may consider
$\C^m\subset\C^n$ as a subspace with projection $P_m$. Then $E$ is
defined by extending each $n\times n$ matrix by zeros for the
additional $(m-n)$ dimensions, and
\begin{equation}
    \label{equation:noiseless1}
        D \mathpunct : \; \varrho \; \mapsto \; P_m \, \varrho \, P_m \, + \,
        \frac{{\rm tr} \, (\1_n - P_m) \varrho}{m} \, P_m\;.
\end{equation}
where the second term serves to make $D$ trace-preserving. Then, clearly,
$DE = \id_m$ as claimed.

To prove the inequality~(\ref{equation:noiseless4}), choose a
maximal family of one-dimensional orthogonal projections $\{ P_\nu
\}_{\nu = 1,...,m} \subset \mathcal{B}_{*}(\C^m)$ such that
$\sum_{\nu = 1}^{m} P_\nu \, = \, \1_m$. Then for any decoding $D
\mathpunct : \; \mathcal{B}_{*}(\C^n) \rightarrow
\mathcal{B}_{*}(\C^m)$, the relation
\begin{equation}
    \label{equation:noiseless2}
        {\rm tr} \, \varrho \, F_\nu \, := \, {\rm tr} \,
        D(\varrho) \, P_\nu \; \; \forall \; \; \varrho \, \in \,
        \mathcal{B}_{*}(\C^n)
\end{equation}
defines a set $\{ F_\nu \}_{\nu = 1,...,m} \subset
\mathcal{B}_{*}(\C^n)$ of positive operators satisfying $\sum_{\nu
= 1}^{m} F_\nu \, = \, \1_n$. For any encoding $E \mathpunct : \;
\mathcal{B}_{*}(\C^m) \rightarrow \mathcal{B}_{*}(\C^n)$ we thus
have
\begin{eqnarray}
    \label{equation:noiseless3}
            n  & = &  {\rm tr} \, \1_n \, = \, {\rm tr} \,
            \sum_{\nu = 1}^{m} \, F_\nu \nonumber \\
            & \geq &  \sum_{\nu = 1}^{m} \, {\rm tr} \, E (P_\nu) \,
            F_\nu \nonumber \\
            & = &
            \sum_{\nu = 1}^{m} \, {\rm tr} \, D \left ( E \left ( P_\nu \right
            ) \right ) \, P_\nu \\
            & \geq &  \sum_{\nu = 1}^{m} \, {\rm tr} \, P_\nu \; -
            \; \sum_{\nu = 1}^{m} \left | \, {\rm tr} \, \left ( P_\nu \,
            D ( E (P_\nu)) \, - \, P_\nu \right ) \, \right |
            \nonumber \\
            & \geq &  m \, - \, \sum_{\nu = 1}^{m} \, \norm{D ( E
            (P_\nu)) \, - \, P_\nu} \nonumber \\
            & \geq &  m \, \left ( 1 \, - \, \cb{D \, E - \id_m}
            \right ) \nonumber,
\end{eqnarray}
where in the third line we have used \Eref{equation:noiseless2}.
\Eref{equation:noiseless3} then immediately implies Equation
(\ref{equation:noiseless4}).

We now have to convert these estimates Equations
(\ref{delta0},\ref{equation:noiseless4}) into statements for
achievable rates for $S=\id_n$ and $T=\id_m$. Thus Equations
(\ref{delta0}) and (\ref{equation:noiseless4}) apply with $n$
replaced by $n^{n_\nu}$ and $m$ replaced by $ m^{m_\mu}$. So let
$(n_\nu)_{\nu \in \N}$ and $(m_\nu)_{\nu \in \N}$ be two integer
sequences such that $\lim_{\nu\to\infty} \, n_\nu \, = \, \infty$
and $\overline{\lim}_{\nu\to\infty} \, \frac{m_\nu}{n_\nu} \, < \,
\frac{\ld n}{\ld m}$. Then for all sufficiently large $\nu$ we
have $n^{n_\nu} \, \geq \, m^{m_\nu}$, and therefore $\Delta(
n_\nu, m_\nu)=0$, which implies that any $R< \frac{\ld n}{\ld m}$
is achievable.

On the other hand, let $R=\frac{\ld n}{\ld m} \, + \,
\varepsilon$, for some $\varepsilon > 0$, and choose diverging
sequences such that $\overline{\lim}_{\nu\to\infty} \,
\frac{m_\nu}{n_\nu} \, =R$. Then $\frac{n^{n_\nu}}{m^{m_\nu}} \,
\leq \, m^{- \varepsilon \, n_\nu}$ infinitely often, and thus, by
Equation (\ref{equation:noiseless4}), $\Delta(n_\nu,m_\nu)$ is
close to $1$ infinitely often. Hence the errors do not go to zero,
and the rate $R$ is not achievable. To summarize, $Q(\id_n,\id_m)
\, = \, \frac{\ld n}{\ld m}$ is the supremum of all achievable
rates. $\blacksquare$\\
\\
By the same techniques, one may also show that the capacity of the
ideal channel does not increase if the information to be
transmitted is restricted to be classical.

\subsection{Partial Transposition Bound}
    \label{sec:ptb}

The upper bound on the capacity of ideal channels can also be
obtained from a general upper bound on quantum capacities, which
has the virtue of being easily calculated in many situations. It
involves, on each system considered, the {\em transposition map},
which we denote by  $\Theta$, defined as matrix transposition with
respect to some fixed orthonormal basis. None of the quantities we
consider will depend on this basis. As is well known,
transposition is positive but not completely positive. Similarly,
we have $\norm\Theta=1$, but generally $\cb\Theta>1$. More
precisely, $\cb\Theta=d$, when the system is described on a
$d$-dimensional Hilbert space \cite{Pau02}. We claim that, for any
channel $T$ and small $\varepsilon > 0$,
\begin{equation}
    \label{equation:cbn-bound}
        Q_{\varepsilon}(T)\leq \ld\Vert T\Theta\Vert_{{\rm cb}} =: Q_\Theta(T),
\end{equation}
where $Q_{\varepsilon} \, ( \geq Q(T) \, )$ is the finite error
capacity introduced in Section \ref{sec:speed}. In particular, for
the ideal channel this implies $Q(\id_d)\leq\ld(d)$.

The proof of Equation (\ref{equation:cbn-bound}) is quite simple
\cite{HW01}: Suppose $R$ is an achievable rate, and that
$\frac{m_\nu}{n_\nu} \rightarrow R\leq Q_\varepsilon(T)$, and
encoding $E_\nu$ and decoding $D_\nu$ are such that
 $\Delta(n_\nu,m_\nu)
  =\cb{D_\nu T^{\otimes n_\nu}E_\nu-\id_2^{\otimes m_\nu}}
  \to0$.
Then we have
\begin{eqnarray}
\eqalign{\fl 2^{m_\nu}
    = \Vert {\rm id}_2^{\otimes
                   m_\nu}\Theta\Vert_{{\rm cb}}
    \leq \Vert ({\rm id}_2^{\otimes m_\nu}
       - D_\nu T^{\otimes n_\nu}E_\nu)\Theta\Vert_{{\rm cb}}
   + \, \Vert D_\nu T^{\otimes
              n_\nu}E_\nu\Theta\Vert_{{\rm cb}}\\
    \lo \leq \Vert \Theta_{2^{m_\nu}}\Vert_{{\rm cb}}\
       \Vert {\rm id}_2^{\otimes m_\nu}-
            D_\nu T^{\otimes n_\nu}E_\nu\Vert_{{\rm cb}}
   + \, \Vert D_\nu (T\Theta)^{\otimes n_\nu}\Theta
             E_\nu\Theta\Vert_{{\rm cb}}\\
    \lo \leq 2^{m_\nu} \, \Delta(n_\nu,m_\nu)
            + \Vert T\Theta\Vert_{{\rm cb}}^{n_\nu},}
\end{eqnarray}
where in the last step we have used that $D_\nu$ and
 $\Theta E_\nu\Theta$ are channels with cb-norm $=1$, and that the cb-norm is exactly tensor
multiplicative, so
 $\Vert X^{\otimes n}\Vert_{{\rm cb}}=\Vert X\Vert_{{\rm cb}}^n$.
Hence, by taking the binary logarithm and dividing by $n_\nu$, we
get
\begin{equation}
    \frac{m_\nu}{n_\nu} +\frac{\ld(1-\Delta(n_\nu,m_\nu))}{n_\nu}
    \leq \ld\Vert T\Theta\Vert_{{\rm cb}}.
\end{equation}
Then in the limit $\nu\to\infty$ we find $R\leq Q_\Theta(T)$ for any
achievable rate $R$. $\blacksquare$\\
\\
The upper bound $Q_{\Theta }(T)$ computed in this way has some remarkable
properties, which make it a capacity-like quantity in its own right. For
example, it is exactly additive:
\begin{equation}
Q_{\Theta }(S\otimes T)=Q_{\Theta }(S)+Q_{\Theta }(T),  \label{capt-add}
\end{equation}
for any pair $S$, $T$ of channels, and satisfies the bottleneck
inequality $Q_{\Theta }(S \, T)\leq \min \{Q_{\Theta
}(S),Q_{\Theta }(T)\}$. Moreover, it coincides with the quantum
capacity on ideal channels: $Q_{\Theta }({\rm id}_{n})=Q({\rm
id}_{n})=\ld n$, and it vanishes whenever $T\Theta $ is completely
positive. In particular, if $\id\otimes T$ maps any entangled
state to a state with positive partial transpose, we have
$Q_{\Theta}(T)=0$.


\section{Alternative Error Criteria}
\label{sec:distances2}

In this section we will show that the various distance measures
introduced in Sections \ref{sec:subspace} through
\ref{sec:entrate} are equivalent, as long as the reference channel
is chosen to be noiseless, i.e., $S = \id_d$ for some $d <
\infty$. Hence the remarkable insensitivity of the quantum
capacity to the choice of the error criterion holds only for the
capacity $Q(T)$ which is our main concern, but not for the more
general $Q(T,S)$ comparing two arbitrary channels. The reason for
this difference is the analogous observation for distance measures
on the state space: different ways of quantifying the distance
between states become inequivalent in the limit of large
dimensions, but all measures for the distance between a {\it pure}
state and a general state essentially agree.


\subsection{Preliminaries}
    \label{sec:prelims}

The following lemma will serve as a starting point for showing the
equivalence of fidelity, ordinary operator norm, and cb-norm
criteria. By $\tracenorm{A} := \trace{\sqrt{A^{*}A}}$ we will
denote the {\em trace norm} of the operator $A \in
\mathcal{B}(\C^d)$, by $\hsnorm{A} := \sqrt{\trace{A^{*}A}}$ its
{\em Hilbert-Schmidt norm}, and by $\norm{A}$ the ordinary
operator norm. These norms are related by the following chain of
inequalities:
\begin{equation}
\label{equation:norms1}
    \norm{A} \, \leq \, \hsnorm{A} \, \leq \, \tracenorm{A}
\end{equation}
(see Chapter VI of \cite{RS80} for a thorough discussion of these
{\em Schatten classes}, these and other useful properties, and the
relation to $\mathcal{L}^p$-spaces). Of course, all norms in a
finite-dimensional space are equivalent, so there must also be a
bound in the reverse direction. This is
\begin{equation}
    \label{equation:norms1a}
        \tracenorm{A} \leq d  \, \norm{A}.
\end{equation}
The crucial difference between these estimates is that the bound
in Equation (\ref{equation:norms1a}) explicitly depends on the
Hilbert space dimension $d$, which makes this inequality useless
for applications in capacity theory, where dimensions grow
exponentially.  Our aim in this section is therefore to relate the
various error measures with dimension-independent bounds (see
Proposition~\ref{propo:distances1} below).

\begin{lemma}
\label{lemma:distances1} Let $\varrho$ be a density operator and
$\psi$ be a unit vector in a Hilbert space $\hh$. Then
\begin{equation}
    \label{equation:distances1}
    \tracenorm{\varrho - \kb{\psi}} \leq 2 \, \sqrt{1 - \bra{\psi}
    \varrho \ket{\psi}},
\end{equation}
with the inequality being strict iff $\varrho$ is pure or $\psi$
is orthogonal to the support of $\varrho$.
\end{lemma}
{\bf {\sc Proof:}} Suppose first that $\varrho = \kb{\varphi}$ is
pure. Then we can compute the trace norm in the two-dimensional
space spanned by $\psi$ and $\varphi$. For the moment we will only
use this property, i.\,e., we assume that $\psi = (1,0)$ is the
first basis vector, and $\varrho$ is an arbitrary $(2 \times 2)$
density matrix. Then we may expand the traceless operator
 $\varrho - \kb{\psi}$ in terms of the Pauli matrices
$\{\sigma_i\}_{i=1,2,3}$, as follows:
\begin{equation}
\label{equation:distances3}
    \varrho - \kb{\psi} = (\varrho_{11}
    -1) \, \sigma_3 \, + \, \re{\varrho_{12}} \, \sigma_1 \, - \,
    \im{\varrho_{12}} \, \sigma_2.
\end{equation}
From this we find the eigenvalues of $\varrho - \kb{\psi}$ to
equal $\pm \sqrt{(\varrho_{11} - 1)^2 + | \varrho_{12} |^2}$, and
hence
\begin{equation}
\label{equation:distances4}
    \tracenorm{\varrho - \kb{\psi}} \, =
    \, 2 \, \sqrt{(\varrho_{11} - 1)^2 + | \varrho_{12} |^2}.
\end{equation}
Positivity of $\varrho$ clearly requires $\det \varrho \geq 0$,
implying $| \varrho_{12} |^2 \leq \varrho_{11} \, \varrho_{22} =
\varrho_{11} \, (1 - \varrho_{11})$. Since $\trace{\varrho^2} = 1
+ 2 \left ( \varrho_{11} (\varrho_{11} - 1) + | \varrho_{12} |^2
\right )$, equality holds if $\varrho$ is pure. Inserting $ |
\varrho_{12} |^2 = \varrho_{11} \, (1 - \varrho_{11})$ into
Equation (\ref{equation:distances4}) directly yields that the
first inequality in Equation (\ref{equation:distances1}) is indeed
strict for pure states.

We now drop the assumption that $\varrho$ is pure and consider an
arbitrary convex decomposition $\varrho = \sum_{i} \lambda_i \,
\varrho_i$ into pure states $\varrho_i$. Then because $x \mapsto
\sqrt{1-x}$ is concave we obtain:
\begin{eqnarray}
    \label{equation:distances5}
        \eqalign{\tracenorm{\varrho - \kb{\psi}} \, & \leq \, \sum_{i} \,
        \lambda_i \, \tracenorm{\varrho_i - \kb{\psi}} \\
        & = \, 2 \, \sum_i \, \lambda_i \sqrt{1 - \bra{\psi}
        \varrho_i \ket{\psi}} \\
        & \leq \, 2 \, \sqrt{1 - \sum_i \, \lambda_i \, \bra{\psi}
        \varrho_i \ket{\psi}} \\
        & = \, 2 \, \sqrt{1 - \bra{\psi} \varrho \ket{\psi}}},
\end{eqnarray}
where in the second step the result for pure states has been used.
This establishes Equation (\ref{equation:distances1}).

Now suppose that equality holds in Equation
(\ref{equation:distances5}). Then because the concavity of $x
\mapsto \sqrt{1-x}$ is strict,  $\bra{\psi} \varrho_i \ket{\psi} =
\bra{\psi} \varrho \ket{\psi} \; \forall \; i$. But since the
convex decomposition of $\varrho$ is arbitrary, we may conclude
that
\begin{equation}
    \label{equation:distances6}
        | \ip{\varphi}{\psi} |^2 \, = \, \bra{\psi} \varrho \ket{\psi} \, \| \varphi \|^2.
\end{equation}
for any vector $\varphi$ in the support of $\varrho$. By
polarization this implies $\ip{\varphi_1}{\psi} \,
\ip{\psi}{\varphi_2} \, = \, \ip{\varphi_1}{\varphi_2} \,
\bra{\psi} \varrho \ket{\psi}$, from which it follows that
\begin{equation}
    \label{equation:distances7}
        S \, \kb{\psi} \, S \, = \, \bra{\psi} \varrho \ket{\psi}
        \, S,
\end{equation}
where $S$ denotes the projection operator on $\supp \varrho$.
Hence either the factor $\bra{\psi} \varrho \ket{\psi}$ vanishes,
entailing $S \psi = 0$, or else $S$ is a rank one operator, and
thus $\varrho$ is pure. This concludes the proof. $\blacksquare$\\
\\
From Lemma \ref{lemma:distances1} we may derive a fidelity-based
expression for the deviation of a given channel from the ideal
channel:
\begin{lemma}
    \label{lemma:distances2}
        Let $\hh$ be a Hilbert space, $\dim \hh < \infty$, and $T \mathpunct : \; \bhstar \rightarrow
        \bhstar$ be a channel. We then have
        \begin{equation}
                \norm{T - \id}  \leq 4 \sup_{\| \psi \| = 1} \left \{ \sqrt{1 -
                \bra{\psi} \, T (\kb{\psi}) \, \ket{\psi}} \right \}.
        \end{equation}
\end{lemma}
{\bf {\sc Proof:}} Note that the operator norm $\norm{T - \id}$
equals the norm of the adjoint operator on the dual space, i.\,e.,
\begin{equation}
    \label{equation:distances8}
        \norm{T - \id} = \sup_{\tracenorm{\varrho} \leq 1}
        \tracenorm{T_{*} \varrho - \varrho}
\end{equation}
(cf. Chapter VI of \cite{RS80} or Section 2.4 of \cite{BR79} for
details). Any matrix $\varrho$, $\tracenorm{\varrho} \leq 1$, has
a decomposition $\varrho = \varrho_1 + i \, \varrho_2$ into
Hermitian $\varrho_i$ satisfying $\tracenorm{\varrho_i} \leq 1$.
Inserting this decomposition into Equation
(\ref{equation:distances8}) and using the triangle inequality, we
find $\norm{T - \id} \leq 2 \, \sup \tracenorm{T_{*} \varrho -
\varrho}$, where the supremum is now over all Hermitian matrices
$\varrho$ obeying $\tracenorm{\varrho} \leq 1$.

By spectral decomposition, any Hermitian matrix $\varrho$ can be
given the form $\varrho = \sum_i r_i \, \varrho_i$, where the
$\varrho_i$ are rank one projectors, and the coefficients $r_i$
are real numbers satisfying $\sum_i | r_i | =
\tracenorm{\varrho}$. Inserting this into the supremum, we see
that $\norm{T - \id} \leq 2 \, \sup \tracenorm{T_{*} \varrho -
\varrho}$, where optimization is now with respect to all
one-dimensional projectors $\kb{\psi}$. The inequality then
directly follows from Lemma \ref{lemma:distances1}. $\blacksquare$


\subsection{Four Equivalent Distance Measures}
    \label{sec:equidist}

We now have in hand all the tools we need to prove that the
distance measures presented in Sections \ref{sec:subspace} through
\ref{sec:entrate} do indeed coincide:
\begin{propo}
    \label{propo:distances1}
        Let $\hh$ be a Hilbert space, $\dim \hh < \infty$, and let $T \mathpunct : \;
        \bhstar \rightarrow \bhstar$ be a channel. Then
        \begin{eqnarray}
                \eqalign{1 \; - \inf_{\varrho \in \bhstar} F_{e}(\varrho,T) & \leq 4 \sqrt{1 -
                F(T)}  \\
                & \leq 4 \sqrt{\norm{T - \id}} \\
                & \leq 4 \sqrt{\cb{T - \id}} \\
                & \leq 8 \left ( 1 - \inf_{\varrho \in \bhstar} F_{e}(\varrho,T) \right
                )^{\frac{1}{4}}.}
        \end{eqnarray}
\end{propo}
These are the dimension independent bounds we need: if a sequence
of channels becomes close to ideal in the sense of any of the
error measures appearing in this proposition, so it will be in
terms of all the others.  The equivalence of the basic capacity
definition \ref{define:basic} based on the cb-norm and the
definitions based on minimum fidelity and entanglement fidelity,
as presented in Sections \ref{sec:subspace} and
\ref{sec:entanglement}, then directly follows.

It is crucial for Proposition~\ref{propo:distances1} that we are
considering only the deviation of $T$ from the ideal channel, so
we can use Lemma~\ref{lemma:distances1} for the distance between
an output state and a pure state. Therefore, for the general
capacity $Q(T,S)$ the choice of the error quantity may remain
important. General properties such as superadditivity
(\ref{equation:superadditive}), which are easy to see for the
cb-norm criterion, might therefore fail for the simpler-looking
operator norm $\norm{T-S}$. This is the principal reason for
choosing the cb-norm in the basic definition.

Mean fidelity, as used in Section~\ref{sec:average1}, and channel
fidelity, as introduced in Section~\ref{sec:entanglement}, are
conspicuously absent from Proposition~\ref{propo:distances1}.
Their role will be discussed in Section~\ref{sec:averagefidel}.

The equivalence of Schumacher's original definition of channel
capacity in terms of the entropy rate will then be treated in
Section~\ref{sec:entmin}.
\\
\\
{\bf Proof of Proposition~\ref{propo:distances1}:} Let $\phi \in
\hh \otimes \hh$ be a purification of $\varrho \in \bhstar$. We
then have
\begin{equation}
\label{equation:distances12}
        1 - F_{e}(\varrho, T)  = \bra{\phi} (\id - T) \otimes \id
        (\kb{\phi}) \ket{\phi}.
\end{equation}
By Schmidt decomposition, $\phi$ can be given a representation
$\ket{\phi} = \sum_j \lambda_j \, \ket{j} \otimes \ket{j'}$, where
$\{ \ket{j} \}_j$ and $\{ \ket{j'} \}_{j}$ are orthonormal systems
in $\hh$, and the so-called Schmidt coefficients $\{ \lambda_{j}
\}_j$ are non-negative real numbers satisfying $\sum_j
\lambda_{j}^2 = 1$. Inserting this representation into Equation
(\ref{equation:distances12}), we see that
\begin{eqnarray}
    \label{equation:distances12a}
        \eqalign{1 - F_{e}(\varrho, T)  & = \sum_{j,k} \lambda_{j}^2 \,
        \lambda_{k}^2 \, \bra{j} (\id - T ) (\op{j}{k}) \ket{k} \\
        & \leq \sum_{j,k} \lambda_{j}^2 \, \lambda_{k}^2 \, \norm{\id
        - T} \; \norm{\op{j}{k}} \\
        & = \norm{\id - T},}
\end{eqnarray}
where in the last step the normalization $\sum_j \lambda_{j}^2 =
1$ has been applied.

The first inequality then immediately follows from Lemma
\ref{lemma:distances2} and the definition of minimum fidelity,
Equation (\ref{equation:subspace}).

An application of the Schwarz inequality directly gives the second
inequality:
\begin{eqnarray}
    \label{equation:distances12b}
        \eqalign{1 - \bra{\psi} \, T (\kb{\psi}) \, \ket{\psi}
        & = \bra{\psi} \, (\id - T) (\kb{\psi}) \, \ket{\psi} \\
        & \leq \norm{\id - T} \; \norm{\kb{\psi}} \\
        & = \norm{T - \id}}
\end{eqnarray}
for all unit vectors $\psi \in \hh$.

The third inequality is obvious from the definition of cb-norm, so we
only need to prove the last step. Applying Lemma \ref{lemma:distances2}
to the operator $T \otimes \id_n$ and then taking the supremum over $n$
on both sides, we see that
\begin{eqnarray}
    \label{equation:distances13}
        \eqalign{\cb{T - \id} & \leq 4 \sqrt{1 - \inf_{n \in \N} F(T
        \otimes \id_n)} \\
        & = 4 \sqrt{1 - \inf_{\varrho \in \bhstar} F_{e}(\varrho,
        T)},}
\end{eqnarray}
concluding the proof. $\blacksquare$

\subsection{Average Fidelity and Channel Fidelity}
    \label{sec:averagefidel}

Average fidelity and channel fidelity have been shown
\cite{HHH99,Nie02} to be directly related error criteria:


\begin{propo}
    \label{propo:knoof}
    Let $\overline{F}(T)$ be the average fidelity and $F_{c}(T)$
    be the channel fidelity of a quantum channel $T \mathpunct : \;
    \bhstar \rightarrow \bhstar$, as introduced in
Equations  (\ref{equation:average}) and
    (\ref{equation:chanfid}), respectively. We then have:
    \begin{equation}
        \label{equation:knoof}
            \overline{F}(T) \, = \, \frac{d \, F_{c}(T) + 1}{d + 1},
    \end{equation}
where $d$ is the dimension of the underlying Hilbert space $\hh$.
\end{propo}
From Proposition \ref{propo:knoof} we may conclude that both
quantities coincide in the large dimension limit $d\to\infty$.
Consequently, average fidelity and channel fidelity are equivalent
error criteria for capacity purposes.

However, neither appears in Proposition~\ref{propo:distances1}.
After giving a somewhat simplified proof of Equation
(\ref{equation:knoof}), we show by an explicit counterexample that
this omission is not accidental. Since a coding for which the
worst case fidelity goes to $1$ also makes the average fidelity go
to $1$, the capacity defined with average fidelity might in
principle be larger than the standard one. That these capacities
nevertheless coincide will then follow from
Proposition~\ref{propo:distances2} in Section \ref{sec:entmin}. A
more direct proof for the equivalence of average fidelity, i.\,e.,
a proof not making use of Proposition \ref{propo:knoof}, is then
presented in Section \ref{sec:average2}. \\
\\
{\bf Proof of Proposition \ref{propo:knoof}:} Suppose that
$\{t_i\}_{i}$ is a set of Kraus operators for the quantum channel
$T$, i.~e., $T(\sigma) = \sum_{i} \, t_{i}^{} \, \sigma \,
t_{i}^{*} \; \; \forall \; \; \sigma \in \bhstar$.

In the course of the proof we will repeatedly employ the so-called
{\em flip} operator $\F \in \bh \otimes \bh$, defined by
$\F(\varphi \otimes \psi) := \psi \otimes \varphi$. In a basis
$\{n\}_{n=1,...,d}$ of $\hh$, $d := \dim \hh$, this corresponds to
the representation
\begin{equation}
    \label{equation:flip}
        \F = \sum_{n,m = 1}^{d} \op{n,m}{m,n}.
\end{equation}
Working in this representation one easily verifies that for all
operators $A, B \in \bh$
\begin{equation}
    \label{equation:magic}
        {\rm tr} \, \F \, (A \otimes B) \, = \, {\rm tr} \, A \, B.
\end{equation}
In terms of the Kraus operators $\{t_i\}_{i}$ the average fidelity
of $T$ then reads
\begin{eqnarray}
    \label{equation:averagefidel1}
            \eqalign{\overline{F}(T) & = \int \bra{U \psi} \, T \, (\kb{U
            \psi}) \, \ket{U \psi} \; dU \\
            & = \sum_i \, {\rm tr} \, \int
            t_{i}^{*} U^{} \varrho \: U^{*} t_{i}^{} U^{} \varrho
            \: U^{*} \; dU \\
            & = \sum_i {\rm tr} \, \F \left ( t_{i}^{*} \otimes
            t_{i}^{} \right ) \int (U \otimes U) \, (\varrho
            \otimes \varrho) \, (U \otimes U)^{*} \; dU,}
\end{eqnarray}
where $\varrho := \kb{\psi}$ is a arbitrary pure reference state,
integration is over all unitaries $U \in \bh$, and in the last
step we have applied Equation (\ref{equation:magic}). The second
factor under the trace,
\begin{equation}
    \label{equation:averagefidel2}
        P(\varrho) := \int (U \otimes U) \, (\varrho
        \otimes \varrho) \, (U \otimes U)^{*} \; dU,
\end{equation}
is obviously invariant under local unitary transformations, i.~e.,
$[P(\varrho) \mid V \otimes V] = 0$ for all unitary operators $V
\in \bh$. Such a state is usually called a {\em Werner state}
\cite{Wer89}, and it follows from the theory of group
representations that these states are spanned by the identity
operator and the flip operator, $P(\varrho) = \alpha \1 + \beta \,
\F$ with complex coefficients $\alpha, \beta$ (see \cite{VW01} and
Chapter~3.1.2 of \cite{Key02} for details). The coefficients can
be easily obtained by tracing $P(\varrho)$ with the identity and
flip operator, respectively, and are both found to equal
$\frac{1}{d(d+1)}$. Inserting the expansion
\begin{equation}
    \label{equation:averagefidel3}
        P(\varrho) = \frac{1}{d(d+1)} \, \left ( \1 + \F \, \right )
\end{equation}
into Equation (\ref{equation:averagefidel1}) and making again use
of Equation (\ref{equation:magic}), we see that
\begin{eqnarray}
    \label{equation:averagefidel4}
        \eqalign{\overline{F}(T) & = \frac{1}{d(d+1)} \sum_i \, {\rm tr} \,
        \F \, (t_{i}^{*} \otimes t_{i}^{}) \, (\1 + \F)\\
        & = \frac{1}{d(d+1)} \left ({\rm tr} \, \sum_i t_{i}^{*} t_{i}^{} \,
        + \, \sum_i {\rm tr} \: t_{i}^{*} \otimes
        t_{i}^{} \right )\\
        & = \frac{1}{d(d+1)} \left ( d + \sum_i \abs{{\rm tr} \,
        t_i}^2 \right )\\
        & = \frac{1}{d(d+1)} \left ( d + d^2 \, F_{c}(T) \right )\\
        & = \frac{1}{d+1} \left ( 1 + d \: F_{c}(T) \right ),}
\end{eqnarray}
where in the second step we have used the normalization $\sum_{i}
\, t_{i}^{*} t_{i}^{} = \1$, and in the third step Equation
(\ref{equation:entanglement24x}) has been applied for the
state $\varrho = \frac{\1}{d}$. $\blacksquare$\\
\\
We proceed with the advertised\\
{\bf Counterexample:} \; For
$\varrho \in \mathcal{B}_{*}(\C^d)$, we set
\begin{equation}
\label{equation:distances24}
    T(\varrho) = P_{+} \varrho P_{+} + P_{-} \varrho P_{-},
\end{equation}
where $P_{+} := \kb{\psi_1}$ is some one-dimensional projector and
$P_{-} := \id - P_{+}$ its ortho-complement. Then by Equation
(\ref{equation:entanglement24x}) we find
\begin{equation}
\label{equation:distances25}
    F_{c}(T) = \frac{1}{d^2} \sum_{i=\pm} | {\rm tr} P_{i} |^2 =
    \frac{d^2 - 2 d + 2}{d^2},
\end{equation}
and therefore $\lim_{d\to\infty} \overline{F}(T) =
\lim_{d\to\infty} F_{c}(T) = 1$, the first equality by
\Eref{equation:knoof}. However, using \Eref{equation:distances8}
we have
\begin{eqnarray}
\label{equation:distances26}
    \norm{T - \id} & = & \sup_{\tracenorm{\varrho} \leq 1}
    \tracenorm{T_{*} \varrho - \varrho} \nonumber \\
    & \geq &  \tracenorm{T_{*} \tilde{\varrho} -
    \tilde{\varrho}} \; \; \forall \; \; \tilde{\varrho} \in
    \mathcal{B}_{*}(\C^d),
\end{eqnarray}
and by choosing $\tilde{\varrho} = \frac{1}{2} \kb{\psi_1 +
\psi_2}$ such that $\psi_2 \perp \psi_{1}$, $\norm{T - \id}$ can
be easily shown to be nonzero, and independent of $d$. Hence there
exists no bound of the form $\norm{T-\id}\leq
f\bigl(F_{c}(T)\bigr)$ with a dimension independent function $f$,
such that $x\to1$ implies $f(x)\to0$.


\subsection{Entanglement Fidelity and Entropy Rate}
\label{sec:entmin}

Let us briefly summarize what we have learned so far about the
interrelation of the various distance measures introduced in
Section~\ref{sec:tema}: From Proposition~\ref{propo:distances1}
and the results of the previous section we may infer the existence
of two classes of equivalent error criteria, one of them
containing average and channel fidelity, the other one cb-norm
distance, operator norm, minimum fidelity, and entanglement
fidelity.

In order to show that both classes lead to the same quantum
channel capacity, we will have to construct, from a given coding
scheme with rate $R$ and channel fidelity approaching one, a
sequence of Hilbert spaces $(\kk_n)_{n\in\N}$ all pure states of
which may be sent reliably with rate $R$. This is the essence of
the following Proposition, which closely follows the argument
presented in Section V of \cite{BKN00}. Although for this purpose
we only need to consider channel fidelities, and thus the chaotic
density operator, the statement is kept more general to apply to
all density matrices, since this will immediately allow us to cope
with Schumacher's definition of channel capacity in terms of
entropy rates as well.
\begin{propo}
\label{propo:distances2}
    Let $\hh$ be a Hilbert space with $d := \dim \hh < \infty$. Let
    $T \mathpunct : \; \bhstar \, \rightarrow \,
    \bhstar$  be a channel, and $\varrho \, \in \, \bhstar$ a density operator.
    Then, for a suitable $k$-dimensional projection $P_k\in\bhh{}$,
    and for the ``compressed channel''
    $T_k \mathpunct : \; \mathcal{B}_{*}(P_k\hh) \, \rightarrow \,
    \mathcal{B}_{*}(P_k\hh)$ given by
  \begin{equation}
        \label{equation:distances22a}
            T_k (\sigma) := P_k \, T(\sigma) \, P_k  +
            {\rm tr}\left( (\1 - P_k)T (\sigma))\right)\,\frac{1}{k}\,P_k,
    \end{equation}
   the estimate
   \begin{equation}
        \label{equation:distances22aa}
            F(P_k\hh,T_k) \, \geq \, 1 \, - \, \frac{1-F_e(\varrho, T)}{1 \, -
            \, q^*},
    \end{equation}
   holds with both
  \begin{eqnarray}
      q^*&=& k\norm{\rho} \qquad \qquad \qquad {\it and}\\
      q^*&=& \frac{1 + \ld d-S(\rho)}{\ld d-\ld k}\;, \label{equation:qstarS}
  \end{eqnarray}
  where $S$ again denotes the von Neumann entropy.
\end{propo}
{\bf {\sc Proof:}} \; The idea of the proof is to recursively
remove dimensions of low fidelity from the support of $\varrho$
until we are left with a Hilbert space of given dimension {\it k}
and a minimum pure state fidelity bounded from below in terms of
$F_e(\varrho,T)$. To this end, define
\begin{equation}
\label{equation:distances16}
    f \mathpunct : \; \hh \rightarrow \R, \hspace{1cm} \ket{\psi} \mapsto \bra{\psi} \, T
    (\kb{\psi}) \, \ket{\psi}.
\end{equation}
Setting $d := \dim \supp{\varrho}$ and $\varrho_0 := \varrho$, we
now recursively define a collection $\{ \varrho_i \}_{i = 0, ...,
d}$ of positive operators, as follows:
\begin{equation}
\label{equation:distances17}
    \varrho_i  :=  \varrho_{i-1}  -  q_i \, \kb{\varphi_i},
\end{equation}
where $\varphi_i$ is the state vector in the support of
$\varrho_{i-1}$ that minimizes {\it f}, and $q_i$ is the largest
positive number that leaves $\varrho_i$ positive. Note that since
$\dim \hh$ is finite, $q_i$ can be chosen to be strictly positive.
By construction, $\supp{\varrho_i}  \subset \supp{\varrho_{i-1}}$,
and $\rank{\varrho_i}  = \rank{\varrho_{i-1}}  -  1$; so our
procedure removes dimensions from the support of $\varrho$ one by
one. It follows that
\begin{equation}
    \label{equation:distances17a}
        \varrho =  \sum_{i=1}^d q_i \, \kb{\varphi_i},
\end{equation}
implying $\sum_{i=1}^d q_i  = \trace{\varrho}  = 1$.

Using the convexity of entanglement fidelity in the density
operator input, we see that
\begin{eqnarray}
    \label{equation:distances19}
        \eqalign{F_e(\varrho, T)  & =  F_e \left ( \sum_{i=1}^d q_i \,
        \kb{\varphi_i},T \right ) \\
        &  \leq  \sum_{i=1}^d  q_i \, f(\varphi_i)\\
        & \leq  f(\varphi_{d-k}) \sum_{i=1}^{d-k}  q_i \; \;
        +  \sum_{i=d-k+1}^{d}  q_i,}
\end{eqnarray}
where in the third line we have used that $k\mapsto f(\varphi_k)$
is non-decreasing by construction. We now take the subspace
$P_k\hh$ as the span of all vectors $\{\varphi_i
\}_{i=d-k+1,\ldots,d}$. Then, since $\bra{\psi} \, T_k (\kb{\psi})
\, \ket{\psi} \geq \bra{\psi} \, T (\kb{\psi}) \, \ket{\psi}$ for
$\psi\in P_k\hh$, we have  $F(P_k\hh,T_k)\geq f(\varphi_{d-k})$.
Introducing $q^*=\sum_{i=d-k+1}^{d} q_i$, and using
$\sum_{i=1}^{d-k}q_i=1-q^*$, we immediately have the desired
estimate.

Our remaining task is to give upper bounds on $q^*$, either in
terms of the largest eigenvalue of $\rho$, or its entropy. Note
that from the sum representing $\varrho$ in
Equation\,(\ref{equation:distances17a}) we have
\begin{equation}
\label{equation:distances18}
    q_i  \leq  q_i  +  \sum_{j=1 \atop j \neq i}^d q_j \: |
    \langle \varphi_j | \varphi_i \rangle |^2  =  \langle
    \varphi_i | \varrho | \varphi_i \rangle  \leq  \norm{\varrho}.
\end{equation}
Therefore, each of the $k$ terms in $q^*$ is bounded by
$\norm{\varrho}$, and we get $q^*\leq k\norm{\varrho}$, which
gives the first estimate.

For the entropic estimate, note first that in the inequality
\begin{equation}
  S(\sum_iq_i\sigma_i)\leq\sum_iq_iS(\sigma_i)-\sum_i q_i \: \ld
  q_i,
\end{equation}
which is valid for arbitrary convex combinations of states
$\sigma_i$ with weights $q_i$ (cf. Chapter 11.3.6 of \cite{NC00}),
the case of pure states $\sigma_i$ leaves just the entropy of the
probability distribution $q$. On the other hand, it is obvious
that among all probability distributions with given weight $q^*$
for the last group of $k$ indices, the one with the highest
entropy is equidistribution, in each of the ranges $1\leq i\leq
d-k$ and $d-k+1\leq i\leq d$. Evaluating the entropy of this
distribution, and combining this with the previous estimate we
find
\begin{eqnarray}
  S(\rho)&\leq&
    H_2 (q^*) +q^* \, \ld k + (1-q^*) \, \ld (d-k)
       \nonumber\\
    &\leq& 1 + \ld d - q^*(\ld d - \ld k),
\end{eqnarray}
where the first term denotes the binary Shannon entropy,
\begin{equation}
    H_2 (q^*) = -q^* \, \ld q^*-(1-q^*) \, \ld(1-q^*)\leq\ 1,
\end{equation}
and we have also used $\ld(d-k)\leq\ld d$. Hence the result
follows by writing this as
an upper bound for $q^*$. $\blacksquare$\\
\\
Proposition \ref{propo:distances2} allows us to make the
transition from average error criteria and entropy rates to
maximal error criteria. So let us assume that a coding scheme
$(E_n,D_n)_{n\in\N}$ is given, together with a sequence
$(\rho_n)_{n\in\N}$ of source states, such that
$F_e(\rho_n,D_nT^{\otimes n}E_n)\to1$. Then the channel $T_k$ will
again be a corrected version of $T^{\otimes n}$, but we can now
conclude that its worst case fidelity goes to one.

Let us first consider the case in which the source does not appear
explicitly, i.e., in which we assume either the mean fidelity or,
equivalently, the channel fidelity to go to one for a scheme with
rate $R$. Since the channel fidelity is just the entanglement
fidelity with respect to a maximally mixed $\rho$, we may apply
Proposition \ref{propo:distances2} with $\rho_n=\1/\dim\hh_n$ and
$\dim\hh_n=2^{\lfloor{nR}\rfloor}$, where conventionally by
$\lfloor x \rfloor$ (read: {\em floor of $x$}) we denote the
largest integer no larger than $x$. We set $k=\dim\hh_n/2$, which
is to say that the modified coding scheme corrects {\it just one
qubit less} than the original one. Then $k\norm{\rho}=1/2$, and we
immediately find that the minimum fidelity is at least
$1-(1-F_e)/2$, and hence also goes to 1.

The second case of interest is that of a source satisfying the
quantum asymptotic equipartition property (QAEP). That is to say,
for large $n$ the Hilbert space $\hh_n$ can be decomposed into a
subspace on which $\rho_n$ essentially looks like a multiple of
the identity, and a subspace of low probability: Given any
$\varepsilon > 0$, for large enough $n$ essentially all the
eigenvalues $\lambda$ of a QAEP quantum source
$(\varrho_n)_{n\in\N}$ with entropy rate $R$ are concentrated in a
so-called {\em $\varepsilon$-typical subspace}, i.\,e.,
\begin{equation}
    \label{equation:qaep}
    2^{-n(R + \epsilon)} \, \leq \, \lambda \, \leq \,
    2^{-n(R - \epsilon)},
\end{equation}
in the sense that the sum of the eigenvalues that do not satisfy
Equation (\ref{equation:qaep}) can be made arbitrarily small. We
can then conclude that $\norm{\rho_n}\leq 2^{-n(R-\varepsilon)}$
for large $n$. Hence, if we choose $k\approx
2^{n(R-2\varepsilon)}$, we can guarantee that $q^*\to0$, and once
again the worst case fidelity has to go to 1. This case covers
product sources and stationary ergodic sources
\cite{BKS+02,BKS+03,BS03}, and many others of interest. The
discussion of the equivalence between the minimum fidelity version
and Schumacher's entanglement fidelity version of channel capacity
in \cite{BKN00} is limited to this case.

Does the equivalence hold even without the equipartition property?
We will give a counterexample below, which is, however, rather
artificial from the point of view of typical coding situations:
the dimension of the spaces $\hh_n$ grows superexponentially. This
is indeed necessary. For if we have an upper bound
$\dim\hh_n\leq\tau^n$ for some positive constant $\tau$,
$S(\rho_n)\approx nR$, and $k\approx 2^{n(R-\varepsilon)}$, we
find that $q^*$ in Equation (\ref{equation:qstarS}) goes to the
constant $(\ld\tau-R)/(\ld\tau-R+\varepsilon)<1$. Therefore, the
maximal subspace fidelity in Equation
(\ref{equation:distances22aa}) in
Proposition~\ref{propo:distances2} goes to one if the
entanglement fidelity does.\\
\\
{\bf Counterexample:} Here we show the claim that the Schumacher
capacity with unconstrained sources is infinite for all channels
with positive quantum capacity. In fact, suppose that we are given
a coding scheme with channel fidelity going to $1$. Then we simply
enlarge the Hilbert space $\hh_n$ by a direct summand $\kk_n$ of
some large dimension, and let
\begin{equation}\label{entropyboosta}
 \rho_n=(1-\varepsilon_n)\frac{\1_{\hh_n}}{\dim\hh_n}
           +\varepsilon_n\frac{\1_{\kk_n}}{\dim\kk_n}
\end{equation}
The coding operations on $\kk_n$ can be completely depolarizing,
for as long as $\varepsilon_n \to0$, the entanglement fidelity of
this source goes to 1, as required. On the other hand, the entropy
of this source is
\begin{equation}\label{entropyboostb}
 S(\rho_n)\geq\varepsilon_n(-\ld\varepsilon_n+ \ld\dim\kk_n)
\end{equation}
Clearly, we can make $S(\rho_n)/n$ diverge if only we let
$\dim\kk_n$ go to $\infty$ fast enough.


\subsection{Entanglement Generation Capacity}
    \label{sec:entgen}

We now focus on Devetak's \cite{Dev03} {\em entanglement
generation capacity}, as introduced in Section
\ref{sec:entanglement}, and verify that it is totally equivalent
to the definitions discussed above. The proof is based on
entanglement-assisted teleportation \cite{BBC93} and therefore
involves classical forward communication from the encoding to the
decoding apparatus. However, this additional resource is shown in
Section \ref{sec:forward} not to affect the quantum channel
capacity.

Due to the additional freedom of choosing an arbitrary pure input
state $\Gamma \in \hh \otimes \hh$ instead of the maximally
entangled state $\Omega$, the entanglement generation capacity is
certainly no smaller than the capacity based on channel fidelity,
which was shown to be a valid figure of merit in the previous
section. So we only need to prove the converse. This is easily
done with the help of teleportation: In the entanglement
generation scenario, the sender and receiver end up sharing a
state $\sigma := (D \, T \, E \otimes \id_{\hh}) \, (\kb{\Gamma})$
which has asymptotically perfect overlap with the maximally
entangled state,
\begin{equation}
    \label{equation:entgen1}
        F := \bra{\Omega} \, \sigma \, \ket{\Omega} \geq 1 - \varepsilon
\end{equation}
for some (small) $\varepsilon> 0$. The output system can thus be
readily interpreted as being in the maximally entangled state with
probability $F \approx 1$, and hence can be used as a resource in
the standard teleportation protocol \cite{BBC93} to transfer
arbitrary quantum states from the sender to the receiver with
fidelity no smaller than $F$, at the same rate $R$.


\section{Isometric Encoding Suffices}
\label{sec:isometric}

In this section we will show that if there exists a coding scheme
that achieves high fidelity transmission for a given source, there
is another coding scheme with isometric encoding, as in Equation
(\ref{equation:isometric}), that also achieves high fidelity
transmission. It then directly follows that in the definition of
channel capacity we may restrict our attention to isometric
encodings, as claimed in Section \ref{sec:restrict}. While this
result is originally due to Barnum et al. \cite{BKN00}, here we
give a slightly generalized version of A.\,S. Holevo's
presentation (cf. Chapter 9 of \cite{Hol02}). All we need is the
following
\begin{propo}
    \label{propo:isometric}
    Let $\hh$, $\kk$ be Hilbert spaces with dimensions $\eta := \dim
    \hh$ and $\kappa := \dim \kk$.
    Let $\varrho \in \bhstar$ be a density
    operator, and $E \mathpunct : \; \bhstar \rightarrow \bkkstar$ a completely positive map
    such that $\trace{E \varrho} = 1$. Let $T \mathpunct : \; \bkkstar \rightarrow \bhstar$ be a
    channel.\\ We may then find a channel $\tilde{E} \mathpunct :
    \; \bhstar \rightarrow \bkkstar$ such that
    \begin{equation}
        \label{equation:isometric1}
        F_e(\varrho, T \, \tilde{E}) \geq (F_{e}(\varrho, T
        \, E))^2,
    \end{equation}
    where for $\sigma \in \bhstar$ we have
    \begin{equation}
        \label{equation:isometric1a}
            \tilde{E}(\sigma) = \left \{
                \begin{array}{l@{\quad : \quad}l}
                    V^{} \, \sigma \, V^{*} & \eta \leq \kappa \\
                    W^{*} \, \sigma \, W^{} + \frac{{\rm tr} \,
                    (\1_{\hh} - W^{} W^{*}) \sigma}{\kappa} \, \1_{\kk} &
                    \eta > \kappa
                \end{array}
            \right.
    \end{equation}
    with isometries $V \mathpunct : \; \hh \rightarrow \kk$
    and $W \mathpunct : \; \kk \rightarrow \hh$, respectively.
\end{propo}
{\bf {\sc Proof:}} \; Let $\{ t_i \}_{i=1,...,\tau}$ and $\{ e_j
\}_{j=1,...,\varepsilon}$ be sets of Kraus operators for the maps
$T$ and $E$, respectively. By Equation
(\ref{equation:entanglement24x}) we have
\begin{equation}
\label{equation:isometric2}
    F_e(\varrho,T \, E) = \sum_{i,j=1}^{\tau, \varepsilon} | {\rm tr} \, t_i \, e_j \, \varrho |^2 =
    \sum_{i,j=1}^{\tau, \varepsilon} | X_{i,j} |^2,
\end{equation}
where $X_{i,j} := {\rm tr} \, t_i \, e_j \, \varrho$. If $\tau
\neq \varepsilon$, add zero components so that $X$ becomes an $(m
\times m)$ square matrix, $m := \max \{ \tau, \varepsilon \}$.

By the singular value decomposition we may find unitary matrices
$A, B$ such that $X = A D B$, where $D$ is diagonal with real
non-negative entries. Since this decomposition simply corresponds
to a change of the Kraus representation of $T$ and $E$, we may
assume without loss that $X$ is diagonal already, and thus
\begin{equation}
\label{equation:isometric3}
    F_e(\varrho,T \, E) = \sum_{i=1}^{m} X_{i,i}^{2} =
    \sum_{i=1}^{m} ({\rm tr} \, t_i \, e_i \, \varrho)^2.
\end{equation}
Now for $k=1,...,m$, we define $\lambda_k := {\rm tr} \, e_k \,
\varrho \, e_{k}^{*}$. Let $\varrho = \sum_j p_j \, \kb{\psi_j}$,
$p_j > 0$, $\sum_j p_j = 1$, be a diagonal representation of the
density operator $\varrho$. Then  $\lambda_k = \sum_j p_j
\norm{e_k \, \psi_j}^2$ and ${\rm tr} \, t_k \, e_k \, \varrho =
\sum_j p_j \, \bra{\psi_j} t_k e_k \ket{\psi_j}$.  Thus,
$\lambda_k = 0$ implies that $e_k \psi_j = 0 \; \forall \; j$, and
therefore ${\rm tr} \, t_k \, e_k \, \varrho = 0$, so that these
terms do not contribute to the sum in Equation
(\ref{equation:isometric3}). We may therefore assume without loss
that $\lambda_k > 0 \; \forall \; k=1,...,m$. Moreover,
\begin{equation}
\label{equation:isometric4}
    \sum_{k=1}^{m} \lambda_k = \sum_{k=1}^{m} {\rm tr} \, e_k \,
    \varrho \, e_{k}^{*} = {\rm tr} \, E(\varrho) = 1,
\end{equation}
where in the last step we have used that $E$ is trace-preserving
on the state $\varrho$. Since
\begin{equation}
\label{equation:isometric5}
    F_e(\varrho,T \, E) = \sum_{i=1}^{m} \lambda_i \, \frac{({\rm
    tr} \, t_i \, e_i \, \varrho)^2}{\lambda_i},
\end{equation}
we may find an index $k$ such that
\begin{equation}
\label{equation:isometric6}
    ({\rm tr} \, t \, e \, \varrho)^2 = \frac{({\rm tr} \, t_k \, e_k \,
    \varrho)^2}{\lambda_k} \geq F_e(\varrho, T \, E),
\end{equation}
where we have introduced the short-hands $e :=
\frac{e_k}{\sqrt{\lambda_k}}$ and $t := t_k$, respectively.\\
Applying the Schwarz inequality we see that
\begin{eqnarray}
    \label{equation:isometric7}
        \eqalign{({\rm tr} \, t \, e \, \varrho)^2 & = \abs{{\rm tr} \, (t^{*} \,
        \sqrt{\varrho})^* \, e \, \sqrt{\varrho} \, }^2 \\
        & \leq \trace{t \, t^* \, \varrho} \; \trace{e^* \, e \, \varrho} \\
        & = \trace{\abs{t^*}^2 \varrho}.}
\end{eqnarray}
Let us treat the case $\eta \leq \kappa$ first: Since $t^* \, t
\leq \1_\kk$, by working in the spectral representation one easily
obtains $t \, t^* \leq \1_\hh$, and $| t^* |^2 \leq | t^* |$, from
which it follows that
\begin{equation}
\label{equation:isometric8}
    {\rm tr} \, | t^* |^2 \varrho \leq {\rm tr} \, | t^* | \varrho
    = {\rm tr} \, t \, V \, \varrho,
\end{equation}
where by $V \mathpunct : \; \hh \rightarrow \kk$ we denote the
polar isometry of $t^{*}$, i.\,e., $t^* = V | t^* |$. Existence of
this isometry requires $\eta \leq \kappa$. Since
\begin{equation}
\label{equation:isometric9}
    \abs{{\rm tr} \, t \, V \, \varrho}^2 \leq \sum_{i=1}^{m}
    \abs{{\rm tr} \, t_i \, V \, \varrho
    }^2 = F_e(\varrho, T \, V(\cdot)V^{*}),
\end{equation}
tracing backwards our results leaves us with the following chain
of inequalities:
\begin{eqnarray}
\label{equation:isometric10}
    F_e(\varrho, T \, V(\cdot)V^{*}) &
    \geq \abs{{\rm tr} \, t \, V \, \varrho}^2
    & {\rm by \;} (\ref{equation:isometric9}) \nonumber\\
    & \geq ({\rm tr} \,
    \abs{t^*}^2 \, \varrho)^2 & {\rm by \;} (\ref{equation:isometric8}) \nonumber\\
    & \geq ({\rm tr} \, t
    \, e \, \varrho)^4 & {\rm by \;} (\ref{equation:isometric7}) \nonumber\\
    & \geq (F_{e}(\varrho,T
    \, E))^2, \qquad & {\rm by \;} (\ref{equation:isometric6})
\end{eqnarray}
which is what we set out to prove.

If $\eta > \kappa$, the proof proceeds very similarly: We denote
the polar isometry of $t$ by $W \mathpunct : \: \kk \rightarrow
\hh$, i.\,e., $t = W \abs{t}$. From Equation
(\ref{equation:isometric7}) we may then conclude that
\begin{eqnarray}
    \label{equation:isometric11}
            \eqalign{({\rm tr} \, t \, e \, \varrho)^2 & \leq {\rm tr} \,
            t^{} \, t^{*} \varrho\\
            & =  {\rm tr} \, W^{} \abs{t}^2 W^{*} \varrho \\
            & \leq  {\rm tr} \, W^{} \abs{t} W^{*} \varrho \\
            & =  {\rm tr} \, t \, W^{*} \varrho,}
\end{eqnarray}
where in the second to last step we have used that $t^{*} \, t
\leq \1_{\kk}$, and thus $\abs{t}^2 \leq \abs{t}$. Substituting
$W^*$ for $V$, we now mimic Equations  (\ref{equation:isometric9})
and (\ref{equation:isometric10}) to conclude that $(F_{e}
(\varrho, T \, E))^2 \leq F_{e} (\varrho, T \, W^{*}(\cdot)W)$.
The map $W^* (\cdot) W$, while being completely positive, is not
necessarily trace-preserving. The desired result then follows by
renormalization, as above in Equations (\ref{equation:noiseless1})
and
(\ref{equation:distances22a}). $\blacksquare$\\
\\
In Proposition \ref{propo:isometric} we have included cases in
which the input space of the channel is strictly smaller than the
input space of the encoding map, $\kappa = \dim \kk < \dim \hh =
\eta$. Though we do not need to consider this situation in our
settings, it may prove helpful in other applications to avoid
cumbersome distinction of cases, and thus has been added for
convenience.

To arrive at the statement that isometric encoding suffices, all
we need to do then is to combine Proposition \ref{propo:isometric}
with the channel fidelity definition of quantum capacity, Section
\ref{sec:entanglement}, taking $\varrho$ to be the chaotic density
matrix, $E$ to be the encoding channel, and thinking of $T$ as the
concatenation of quantum channel and decoding channel.


\section{Classical Side Information}
    \label{sec:sideinfo}

As claimed in Section \ref{sec:friend}, it is a straightforward
consequence of Proposition \ref{propo:isometric} that classical
forward communication has no effect on the quantum channel
capacity \cite{BDS+96,BKN00}. However, before entering the proof
we need to make a few more comments on the assisted channel $T
\otimes \id^{c}_{\Lambda}$, where $T \mathpunct : \; \bhhstar{1}
\rightarrow \bhhstar{2}$ is an arbitrary quantum channel and
$\id^{c}_{\Lambda}$ denotes the identity on a classical system
with $\Lambda \in \N$ states. Thus, in the limit of large
dimensions an $n$-fold tensor product of the channel $T$ will be
assisted by a classical system with a total of $\Lambda^n$ states.

The results we are going to present in this section apply slightly
more generally: Instead of {\em a priori} fixing a side channel of
given (if arbitrarily large) dimension, the encoder may choose the
size of the side channel in the encoding process, which also
covers the case of super-exponentially growing side channels. The
capacity of a channel $T$ assisted by this type of classical
forward communication will be marked with a subscript,
$Q_{cf}(T)$. This generalization will play a role in Section
\ref{sec:average2}.

Of course, $Q_{cf}(T) \geq Q(T \otimes \id^{c}_{\Lambda}) \geq
Q(T)$. It is the aim of Section \ref{sec:forward} to show that all
these capacities are equal.


\subsection{Classical Forward Communication does not Increase the
Channel Capacity}
    \label{sec:forward}

Obviously, for classically assisted channels the encoding is a
channel with both a classical and a quantum output. Such channels
are usually called {\em instruments} \cite{Dav76}, and can be
thought of as a collection of trace-nonincreasing operators $\{
E_{\lambda} \}_{\lambda = 1,...,\Lambda}$ summing up to a channel
$E = \sum_{\lambda = 1}^{\Lambda} E_\lambda$. The index $\lambda
\in \{ 1,...,\Lambda \}$ represents classical information that may
be obtained in the encoding process and sent undisturbed to the
decoder over a noiseless classical channel. Depending on the value
of $\lambda$, one channel $D_{\lambda_0}$ out of a collection of
trace-preserving quantum channels $\{ D_\lambda
\}_{\lambda=1,...,\Lambda}$ is used in the decoding process.

The definition of achievable rates and channel capacity now
completely parallels the definition of the unassisted quantities.
Here we focus on the channel fidelity version of channel capacity
(cf. Section \ref{sec:entanglement}), since this is a definition
Proposition \ref{propo:isometric} is well suited for. Of course,
all other definitions of channel capacity can be extended to
classically assisted capacities in the same spirit.

We may thus say that $R$ is an achievable rate for the classically
assisted quantum channel $T$ iff there is a sequence of Hilbert
spaces $(\kk_n)_{n \in \N}$ satisfying
$\overline{\lim}_{n\to\infty} \frac{\ld \dim \kk_n}{n} = R$ and a
sequence of encodings $(E_{\lambda_n,n})_{\lambda_n =
1,...,\Lambda_n,n\in\N}$ and decodings
$(D_{\lambda_n,n})_{\lambda_n = 1,...,\Lambda_n,n\in\N}$ for some
integer sequence $(\Lambda_n)_{n\in\N}$ such that
\begin{equation}
    \lim_{n\to\infty} \, \sum_{\lambda_n=1}^{\Lambda_n}
    F_{c} (D_{\lambda_n,n} \, T^{\otimes n} \, E_{\lambda_n,n})
    = 1.
\end{equation}
The quantum capacity $Q_{cf}(T)$ of the channel $T$ with classical
forward communication is then defined as the supremum of all
achievable rates.

Of course, the capacity of the channel $T \otimes
\id^{c}_{\Lambda}$ for fixed $\Lambda \in \N$ is obtained by
setting $\Lambda_n := \Lambda^n$ in the above definition.
\begin{theo}
    \label{theo:forward}
    Let $T \mathpunct : \; \bhhstar{1} \,\longrightarrow \, \bhhstar{2}$ be a quantum
    channel and $\Lambda \in \N$. We then have
    \begin{equation}
        Q_{cf}(T) = Q(T \otimes \id^{c}_{\Lambda}) = Q(T)
    \end{equation}
\end{theo}
Before giving the proof let us consider a seeming generalization
of this theorem, which allows the side channel $R$ to be any {\it
separable channel}, i.e., a channel $R=R_2R_1$ operating by first
collecting classical information, by a channel $R_1$, say and then
recoding this into quantum information by another channel $R_2$.
Equivalently, $(\id\otimes R)$ maps any input state to a separable
state, so these channels are also called ``entanglement breaking''
\cite{HSR03,Rus03}. Then, for every such $R$ and any channel $T$
we have the following
\begin{coro}
    \label{coro:sepchan}
    Separable side channels do not increase the quantum channel
    capacity, i.\,e., for any quantum channel $T$ and any separable
    channel $R = R_2 \circ R_1$ we have
    \begin{equation}
        \label{e:sidechan}
            Q(T\otimes R)=Q(T)\;.
    \end{equation}
\end{coro}
{\bf Proof of Corollary \ref{coro:sepchan}:} \; For a separable
channel $R=R_1 \circ R_2=R_2 \circ \id^{c}_\Lambda \circ R_1$, we
have
\begin{equation}
    \label{e:sepside}
    \eqalign{Q(T)&\leq Q(T\otimes R)
        =Q\bigl((\id\otimes R_2)(T\otimes\id^{c}_\Lambda)
                (\id\otimes R_1)\bigr)\\
        &\leq Q(T\otimes\id^{c}_\Lambda)
        =Q(T)\;,}
\end{equation}
where the first inequality follows by using codings ignoring the
channel $R$, the second follows by the bottleneck inequality
(\ref{equation:bottleneck}), and in the last step we have applied
Theorem \ref{theo:forward}. From this chain of inequalities we get
$Q(T)= Q(T\otimes R)$, just as claimed. $\blacksquare$\\
\\
We now follow \cite{BKN00} in the\\
{\bf Proof of Theorem \ref{theo:forward}:} \; From our remarks it
is clear that we only need to prove the inequality $Q_{cf}(T) \leq
Q(T)$. Given a sequence of Hilbert spaces $(\kk_n)_{n \in \N}$ and
suitable encodings $(E_{\lambda_n,n})_{\lambda_n =
1,...,\Lambda_n,n\in\N}$ and decodings
$(D_{\lambda_n,n})_{\lambda_n = 1,...,\Lambda_n,n\in\N}$ such that
the classically assisted channel $T$ achieves the rate $R$ with
channel fidelity approaching one, we will show that the same rate
can be achieved without using the side channel.

From the definition of channel capacity, for any $\varepsilon > 0$
we may find $n_\varepsilon \in \N$ such that
\begin{equation}
\label{equation:forward1}
    \sum_{\lambda_n=1}^{\Lambda_n}
    F_{c}(D_{\lambda_n,n} \, T^{\otimes n} \, E_{\lambda_n,n})
    \, \geq \, 1 - \varepsilon \; \; \; \forall \; \; n \geq n_\varepsilon.
\end{equation}
For the remainder of the proof we will fix $n \geq n_\varepsilon$
and drop the index to streamline the notation. Setting $e_\lambda
:= {\rm tr \,} E_\lambda \1 / d$ with $d := \dim \kk_n$ and
\begin{equation}
\label{equation:forward2}
    \tilde{F}_{c}(D_{\lambda} \, T \, E_{\lambda}) \, := \,
    \frac{1}{e_\lambda} \,
    F_{c}(D_{\lambda} \, T \, E_{\lambda}),
\end{equation}
we may then rewrite Equation (\ref{equation:forward1}) as
\begin{equation}
\label{equation:forward3}
    \sum_{\lambda=1}^{\Lambda} e_\lambda \;
    \tilde{F}_{c}(D_{\lambda} \, T \, E_{\lambda}) \, \geq
    \, 1 - \varepsilon.
\end{equation}
Since $\sum_{\lambda=1}^{\Lambda} e_\lambda = 1$, there is an
index $\mu$ such that $\tilde{F}_{c}(D_{\mu} \, T \, E_{\mu}) \,
\geq \, 1 - \varepsilon.$ However,
\begin{equation}
\label{equation:forward4}
    \tilde{F}_{c}(D_\mu \, T \, E_\mu) \, = \,
    F_{c} \left(D_\mu \, T \, \frac{E_\mu}{e_\mu} \right ),
\end{equation}
and we may therefore apply Proposition \ref{propo:isometric} to
conclude that there is a channel $\tilde{E}$ such that
\begin{equation}
\label{equation:forward5}
    F_{c}(D_\mu \, T \, \tilde{E}) \geq (1 -
    \varepsilon)^2,
\end{equation}
from which it follows that the same rate can be achieved without
relying on the classical side channel. $\blacksquare$


\subsection{Average Fidelity by Forward Communication}
    \label{sec:average2}

We already know that average fidelity may be considered a suitable
error criterion for capacity purposes. The line of thought we
followed to establish this fact proceeded via the equivalence of
average fidelity and channel fidelity, Proposition
\ref{propo:knoof}, and is thus ultimately based on Proposition
\ref{propo:distances2}.

The results of the previous section regarding the uselessness of
classical forward communication can be employed to give an
alternative proof that average fidelity serves as a valid distance
measure, making use only of the sufficiency of isometric encoding.

To this end, we will show that instead of evaluating the average
fidelity $\overline{F}(T)$ for a given channel $T$, we may just as
well compute the minimum fidelity $F(\tilde{T})$, where the new
channel $\tilde{T}$ is simply the old channel $T$ augmented by
classical forward communication. However, by Theorem
\ref{theo:forward} the quantum capacities of $\tilde{T}$ and $T$
coincide, and therefore average fidelity and minimum fidelity turn
out to be equivalent error quantities.

While our concept of classically assisted channel capacity, as
presented in Section \ref{sec:forward}, only allows for discrete
classical messages and therefore involves the calculation of
finite sums, for the evaluation of average fidelity we will rather
deal with integrals, corresponding to continuous classical
messages. However, this extension poses no difficulties.\\
\\
So suppose we have at hand a quantum channel $T \mathpunct : \;
\bhhstar{1} \rightarrow \bhhstar{2}$ together with encoding
channel $E \mathpunct : \; \bhhstar{3} \rightarrow \bhhstar{1}$
and decoding channel $D \mathpunct : \; \bhhstar{2} \rightarrow
\bhhstar{3}$. By Equation (\ref{equation:average}) the average
fidelity of the concatenation $D \circ T \circ E$ then reads
\begin{eqnarray}
    \label{equation:average1}
        \eqalign{\overline{F}(\hh_3, D \, T \, E) &
        = \int \bra{U \psi} D \, T \, E \, (\kb{U
        \psi}) \, \ket{U \psi} \; dU \\
        & = {\rm tr} \, \left ( \varrho \int U^* D \, T \, E \, (U \varrho
        U^*) \, U \; dU \right ),}
\end{eqnarray}
where integration is over all unitaries $U \in \bhhstar{3}$, and
$\varrho = \kb{\psi} \in \bhhstar{3}$ is an arbitrary reference
state. We will now convert $D \, T \, E$ into a channel with
classical forward communication. Denoting by $\mathcal{C}(X)$ the
vector space of continuous functions on the set $X$, we may define
$\tilde{D} \mathpunct : \; \bhhstar{2} \otimes
\mathcal{C}(U(\hh_3)) \; \rightarrow \; \bhhstar{3}$ and
$\tilde{E} \mathpunct : \; \bhhstar{3} \; \rightarrow \;
\bhhstar{1} \otimes \mathcal{C}(U(\hh_3))$, as follows:
\begin{eqnarray}
    \label{equation:average2}
        \eqalign{&\tilde{D} (\varrho \otimes f) := \int U^* D(\varrho) \, U \,
        f(U) \; dU, \\
        & \tilde{E}_{U} (\sigma) := E(U \sigma U^*),}
\end{eqnarray}
where in the definition of $\tilde{E}$ we have made use of the
fact that $\bh \otimes \mathcal{C}(X)$ is isomorphic to the
$\bh$-valued functions on $X$. We then see that for any state
$\varrho = \kb{\psi} \in \bhhstar{3}$
\begin{eqnarray}
    \label{equation:average3}
       \eqalign{\bra{\psi} \tilde{D} \, (T \otimes &
       \id_{\mathcal{C}(U(\hh_3))}) \, \tilde{E} \, (\kb{\psi})
       \, \ket{\psi} \\
       & = {\rm tr} \, \varrho \, \tilde{D} \, (T \otimes
       \id_{\mathcal{C}(U(\hh_3))}) \, \tilde{E} (\varrho) \\
       & = {\rm tr} \, \left ( \varrho \int U^* D \, T \, E \, (U \varrho
        U^*) \, U \; dU \right ).}
\end{eqnarray}
The average fidelity for the concatenation $D \, T \, E$ thus
equals the minimum pure state fidelity for the classically
assisted channel $\tilde{D} \, (T \otimes \id) \, \tilde{E}$,
which is what we wanted to show.

Note that for this proof to apply also in the setting of $n$-fold
tensor products, as required by the definition of channel
capacity, we may not restrict ourselves to exponentially growing
side channels, $(T \otimes \id_{\Lambda}^{c})^{\otimes n}$: This
would correspond to an averaging over $n$-fold tensor products of
unitary operators, $U_1 \otimes U_2 \otimes \cdots \otimes U_n$.
However, not all unitary operators on an $n$-fold tensor product
are of this form.


\section{Testing Only One Sequence}
    \label{sec:sequences2}

We will now prove the claim made in Section~\ref{sec:sequences1}:
If a coding scheme construction works for a certain pair of
integer sequences $(n_\nu)_{\nu\in\N}$, $(m_\nu)_{\nu\in\N}$ such
that the rate $R$ is achieved infinitely often, i.\,e.,
$\overline{\lim}_{\nu\to\infty} \frac{m_\nu}{n_\nu} = R$, and the
error tends to zero, $\lim_{\nu\to\infty} \Delta (n_\nu, m_\nu) =
0$, then coding works for all such pairs.

As mentioned in Section~\ref{sec:sequences1}, this requires
extending a given coding scheme to more block sizes. Therefore
this section will be organized by extension method: In
Section~\ref{sec:sequences2a} we use only the method of {\it
wasting resources}, i.e., either using the coded channels for
fewer bits than allowed by the given coding scheme (i.e.,
decreasing $m_\nu$) or requiring some additional channel uses
(thus increasing $n_\nu$) and simply not using them. This will
allow the extension whenever we can find a subsequence along
which, on the one hand, the desired rate is achieved and which, on
the other hand,  does not grow too fast.

A second method would be to use blocks from the given coding
scheme and put them together as tensor products, to get to larger
block sizes. We show in Section~\ref{sec:sequences2b} by an
explicit example that this method, combined with the wasteful one,
is not sufficient to extend a very sparse coding sequence to all
large block sizes.

Finally, we show in Section~\ref{sec:sequences2c}, based on the
work in \cite{KW02}, how hashing codes can be used to achieve the
desired extension in all cases.

Throughout, we will denote by
$(N_\mu)_{\mu\in\N},(M_\mu)_{\mu\in\N}$ the given coding
sequences, and assume, without loss of generality, that the
sporadic rate is attained, i.e., $R=\overline{\lim}_{\mu\to\infty}
\frac{M_\mu}{N_\mu}=\lim_{\mu\to\infty} \frac{M_\mu}{N_\mu}$. The
sequences to which we seek to extend the scheme are denoted by
$(n_\nu)_{\nu\in\N}$ and $(m_\nu)_{\nu\in\N}$, as before.


\subsection{Subexponential Sequences}
\label{sec:sequences2a}

Obviously, good coding becomes easier the more parallel channels
are available for transmission. Moreover, if a certain coding
scheme works for some Hilbert space $\hh$, it works at least as
well for states supported on a lower dimensional Hilbert space
$\hh'$. Thus, the error quantity $\Delta (n,m)$, as introduced in
Definition~\ref{define:basic}, has the following monotonicity
properties:
\begin{equation}
\label{equation:mono}
    \Delta (n+1,m) \leq \Delta(n,m) \leq \Delta(n,m+1)
\end{equation}
for all positive integers $n,m$. We call a  diverging sequence
$(N_\mu)_{\mu\in\N}$ {\em subexponential} if
\begin{equation}\label{e:subex}
  \lim_{\mu\to\infty} \frac{N_{\mu+1}}{N_\mu} = 1\;.
\end{equation}
This covers, for example, all arithmetic sequences, and
polynomially growing ones. For such sequences the desired result
follows directly from the following Lemma, which slightly
generalizes Lemma 3.2 of \cite{KW02}.
\begin{lemma}
\label{lemma:sequences}
    Suppose $\Delta \mathpunct : \; \N \times \N \rightarrow
    \R_{+}$ satisfies the monotonicity properties
    (\ref{equation:mono}). Let $(N_{\mu})_{\mu\in\N}$,
    $(M_{\mu})_{\mu\in\N}$ be a pair of integer sequences such that
    $(N_\mu)_{\mu\in\N}$ is subexponential, cf. Equation (\ref{e:subex}),
    and  $\lim_{\mu\to\infty}
    \Delta(N_\mu,M_\mu) = 0$.\\ Then for any pair of integer
    sequences $(n_{\nu})_{\nu\in\N}$, $(m_{\nu})_{\nu\in\N}$ such
    that $\lim_{\nu\to\infty} n_\nu = \infty$ and
    \begin{equation*}
        \label{equation:dense0}
        \overline{\lim}_{\nu\to\infty} \frac{m_\nu}{n_\nu} <
        \underline{\lim}_{\mu\to\infty} \frac{M_\mu}{N_\mu},
    \end{equation*}
    we have $\lim_{\nu\to\infty} \Delta(n_\nu,m_\nu) = 0$.
\end{lemma}
{\bf {\sc Proof:}} \; If we have only the monotonicity property of
$\Delta$ to draw upon, the way to show that
$\Delta(n_\nu,m_\nu)\to0$ is to find a suitable index
$\mu=\mu(\nu)$ for all sufficiently large $\nu$ so that
$\Delta(n_\nu,m_\nu)\leq\Delta(N_{\mu(\nu)},M_{\mu(\nu)})$, for
which we need
\begin{equation}\label{e:nnuNmu}
  n_\nu\geq N_{\mu(\nu)}  \quad \mbox{ and}\quad
  m_\nu\leq M_{\mu(\nu)}\;.
\end{equation}
The first inequality we will ensure by defining
\begin{equation}
    \label{equation:dense1}
    \mu({\nu}) = \min \{ \alpha \; | \; N_\alpha \geq n_\nu \} -
    1.
\end{equation}
Then
\begin{equation}
    \label{equation:dense2}
    N_{\mu(\nu)} \leq n_\nu \leq N_{\mu({\nu})+1} \; ,
\end{equation}
and $\lim_\nu \mu(\nu)=\infty$. Hence it remains to show that the
second inequality in Equation\,(\ref{e:nnuNmu}) holds for all
sufficiently large $\nu$. We consider
\begin{equation}\label{e:quotients}
  \frac{m_\nu}{M_{\mu(\nu)}}
    =\frac{m_\nu}{n_\nu}\;  \frac{n_\nu}{N_{\mu(\nu)+1}}\;
      \frac{N_{\mu(\nu)+1}}{N_{\mu(\nu)}}\;
      \frac{N_{\mu(\nu)}}{M_{\mu(\nu)}}.
\end{equation}
In this product the second factor is $\leq1$ by
Equation\,(\ref{equation:dense2}), and the third converges to $1$
because $(N_\mu)_{\mu\in\N}$ is subexponential. Now pick $R_-,R_+$
such that strict inequalities
\begin{equation}\label{equation:dense0R}
        \overline{\lim}_{\nu\to\infty} \frac{m_\nu}{n_\nu} < R_- <R_+<
        \underline{\lim}_{\mu\to\infty} \frac{M_\mu}{N_\mu},
\end{equation}
hold. Then for all sufficiently large $\nu$ the first factor in
Equation\,(\ref{e:quotients}) is $\leq R_-$, and the last factor
is $\leq1/R_+$. Hence the product of the first and last factor in
Equation\,(\ref{e:quotients}) is  $\leq R_-/R_+<1$. Hence
Equation\,(\ref{e:nnuNmu})
holds for all sufficiently large $\nu$. $\blacksquare$\\

This result covers most sequences $(N_\mu)_{\mu\in\N},
(M_\mu)_{\mu\in\N}$ naturally arising for families of codes. In
contrast to Proposition~\ref{propo:hashing1} the result therefore
remains useful even for the simulation of one noisy channel by
another, i.e., for the definition of capacities $Q(T,S)$ with
non-ideal reference channel $S$.

\subsection{A Counterexample}
\label{sec:sequences2b}

From the way in which subexponentiality of $(N_\mu)_{\mu\in\N}$
enters the proof of Lemma~\ref{lemma:sequences}, it is not clear
whether this assumption is really necessary. In this section we
will give an example showing that it cannot be omitted, implying
that to establish full equivalence we do need the more
sophisticated techniques presented in
Section~\ref{sec:sequences2c}. The example will also satisfy
another natural constraint on the error function $\Delta(n,m)$,
which reflects another elementary method of getting new coding
schemes from old: we can always split the given number of channels
into sub-blocks, and apply a known coding scheme to each block.
The total error is then estimated as the sum of the errors of each
block. Hence the error function $\Delta(n,m)$ is {\it
subadditive}:
\begin{equation}
    \label{equation:sub}
    \Delta(n_1 + n_2, m_1 + m_2) \leq \Delta(n_1,m_1) +
    \Delta(n_2,m_2).
\end{equation}
Suppose we are given some codes for a possibly very sparse
sequence of block sizes $N_\mu$, with $M_\mu=N_\mu$ coded bits,
and $\varepsilon_\mu=\Delta(N_\mu,N_\mu)\to0$. Then the rate $1$
is sporadically achievable. The error bound we get by the best
combination of blocking and possibly wasting some resources is
then
\begin{equation}
    \label{equation:countex1}
    \Delta(n,m) := \inf \left \{ \sum_k \varepsilon_{\mu_k} \; | \; m
    \leq \sum_k N_{\mu_k} \leq n \right \}\;,
\end{equation}
where the infimum is taken over all admissible sets
$\{N_{\mu_k}\}$. This satisfies both monotonicity
(\ref{equation:mono}) and subadditivity (\ref{equation:sub}). Our
aim in constructing the counterexample is to choose
$(N_\mu)_{\mu\in\N}$ growing sufficiently rapidly, and
$(\varepsilon_\mu)_{\mu\in\N}$ decreasing sufficiently slowly, so
that $\Delta(n,m)$ can be bounded away from zero even though $m/n$
gets small.

We assume that $(N_\mu)_{\mu\in\N}$ is {\it superexponential} in
the sense that $N_{\mu+1}/N_\mu\to\infty$. Of
$(\varepsilon_\mu)_{\mu\in\N}$ for the moment we only require that
it decreases monotonically to zero. Then the infimum in
Equation\,(\ref{equation:countex1}) never contains sums arising by
breaking up a block of size $N_{\mu_k}$ into blocks of smaller
sizes: in this way one would not only get more terms in the sum of
$\varepsilon$'s, but each term would be larger than
$\varepsilon_{\mu_k}$. Therefore we can lower bound $\Delta$ by
considering only a decomposition into the largest available
blocks. For $N_\mu\leq m\leq n<N_{\mu+1}$ this means
\begin{equation}\label{e:countex137}
  \Delta(n,m)\geq \varepsilon_\mu\;\left\lfloor \frac m{N_\mu}
  \right\rfloor\;,
\end{equation}
where $\lfloor x\rfloor$ denotes the largest integer $\leq x$. Now
we choose $n_\mu=N_{\mu+1}-1$, and $m_\mu$ close to the geometric
mean: $m_\mu \approx \sqrt{N_\mu N_{\mu+1}}$. Then on the one hand
$m_\mu/n_\mu\approx \sqrt{N_\mu /N_{\mu+1}}\to0$, because
$(N_\mu)_{\mu\in\N}$ is superexponential. Hence this pair of
sequences has rate zero. On the other hand, if we only let
$\varepsilon_\mu$ decrease slowly enough we can prevent
$\Delta(n_\mu,m_\mu)$ from going to zero. For example, with
$\varepsilon_\mu=\sqrt{N_{\mu}/N_{\mu+1}}/2$ we get
$\Delta(n_\mu,m_\mu)\geq1/2$ asymptotically.

To summarize: we have constructed a monotone and subadditive
function $\Delta(n,m)$, for which the rate $1$ can be achieved
sporadically, but for which the proper achievable rate is $0$.


\subsection{Hashing Helps}
\label{sec:sequences2c}

We will now explain how hashing helps to establish full
equivalence of the one-sequence and all-sequence definitions,
showing that it is indeed sufficient to check only one pair of
sequences when testing a given rate $R$. As we know from the
previous subsection, this requires that if we have found a fairly
good coding for some large block size, we must make better use of
it than just repeating the blocks, and maybe not using some of the
input bits.

This problem is in essence the same that arises when we have a
fairly good channel to begin with: just repeating it without
further encoding will not make errors go to zero. Instead they
will accumulate. But, on the other hand, a channel which is nearly
ideal should also have nearly the capacity of an ideal channel, or
else the whole idea of capacity would make no sense. In fact, in
our paper so far we have only shown one type of channel to have
positive capacity, namely the ideal channels. As shown in Section
\ref{sec:ideal2}, in that case the problem of accumulating errors
simply does not occur, and all coding is with $\Delta(n,m)=0$.

So it would actually be conceivable that capacities are always
zero, unless coding can be done without errors (see also
Section~\ref{sec:speed}). However, it can be shown that {\it small
errors can be corrected} with only a small loss in capacity. This
problem is treated in a self-contained way in \cite{KW02}, and we
refer to that paper for details and proofs. Here we only point out
the statements needed in the present context, and sketch the main
arguments.

The non-trivial family of codes needed for this argument are
called hash codes. In \cite{KW02} they are constructed as {\it
random graph codes}, based on a scheme \cite{SW99} which turns
graphs into quantum error correcting codes of the Knill-Laflamme
type. The verification that a certain number of errors is
corrected by such a code amounts to showing that a certain system
of linear equations is non-singular. Then the existence of codes
with suitable parameters is shown by checking that this condition
holds true in a generic random graph of suitable size. The random
graphs are generated such that the probabilities for each edge are
independent and equidistributed. This is quite different from
Shannon's idea of random coding, where the distribution depends on
the noise in the channel and the input state. The argument based
on graph codes works in any Hilbert space dimension $d$ which is a
prime number. It shows that if we want to encode $m$ systems of
dimension $d$ into $n$ systems of the same dimension, and
\begin{equation}\label{e:randgraph}
  \Bigr(\frac mn+\frac{4f}n-1\Bigl)\ld d
        +H_2\Bigl(\frac{2f}n\Bigr) <0
\end{equation}
then we can arrange for the code to correct arbitrary errors
occurring on up to $f$ of the $n$ subsystems. Here
\begin{equation}\label{e:binent}
  H_2(p)=-p\;\ld(p)-(1-p)\,\ld(1-p)
\end{equation}
is again the binary entropy function. Moreover, the expression in
Equation (\ref{e:randgraph}) is an upper bound on the exponential
rate at which the probability for a random graph code {\it not} to
correct that many errors decreases. The crucial feature is that if
$f/n$ is small, i.e., we do not require many errors to be
corrected, then we can get $m$ close to $n$, i.e., the rate of the
coding scheme is nearly that of the ideal channel.

The next step is to convert the correction of {\it rare} errors to
that of arbitrary {\it small} errors. Here a straightforward norm
estimate is
\begin{equation}\label{e:raresmall}
  \cb{DT^{\otimes n}E-\id_m}
   \leq\left(2^{H_2((f+1)/n)}\; \cb{T-\id}^{\frac{f+1}{n}}\right)^n\;,
\end{equation}
when $E,D$ are a code correcting $f$ out of $n$ errors on $m$
input systems, as above, and $\id_m$ denotes the ideal channel on
$m$ $d$-level systems. Then as soon as the expression in
parentheses is $<1$, in particular, if the channel $T$ is close to
ideal, we see that the errors go to zero exponentially in $n$.

We now apply these ideas to a given coding solution for some
channel $T$, i.e., we assume that for the given channel we have
some encoding of a $d$-level system through $N$ parallel uses of
the channel. The nominal rate of this coding scheme, expressed in
the units ``qubits per channel use'' is $\ld d/N$. For large $N$
we may as well assume that $d$ is a prime number, because the gaps
between consecutive primes go to zero \cite{Ing37}. Then we apply
the above ideas to the encoded channel $\widetilde T=D(T^{\otimes
N})E$. The overall code will require $nN$ channel uses, and the
encoded systems are $d^m$ dimensional. If we use $n$ as the index
of the resulting sequence of channel uses, the resulting sequence
of block lengths grows linearly, hence is clearly subexponential,
and has rate $(m \, \ld d)/(nN)$. The errors go to zero
exponentially, provided we can find $f$ satisfying Equation
(\ref{e:randgraph}) and such that the parenthesis in Equation
(\ref{e:raresmall}) is strictly less than one. Combining all this
gives the following estimate (Theorem~8.2 in \cite{KW02}):
\begin{propo} \label{propo:hashing1}
Let $T$ be a channel, not necessarily between systems of the same
dimension. Let $N,d\in{\mathbb N}$ with $d$ a prime number, and
suppose that there are channels $E$ and $D$ encoding and decoding
a $d$-level system through $N$ parallel uses of $T$, with error
 $\Delta=\cb{DT^{\otimes N}E-\id_d} < \frac{1}{2e}$, with $e=\exp1$. Then
\begin{equation}\label{e:Qforall}
  Q(T)\geq \frac{\ld d}N(1-4e\Delta)-\frac1NH_2(2e\Delta)\;.
\end{equation}
Moreover, $Q(T)$ is the least upper bound on all expressions of
this form, and for coding rates below the bound the errors
decrease exponentially.
\end{propo}
Note that here a single successful coding scheme $(E,D)$
guarantees at least a lower bound to the capacity. The most
important aspect of this bound is once again that the precision
$\Delta$ required does not depend on the dimension $d$. Therefore,
even if we know such codes only on an arbitrarily thinly spaced
sequence of $N$'s, with vanishing errors along this thin
subsequence, we can achieve all rates below the sporadic rate
($\overline{\lim}_\mu \ld d_\mu/N_\mu$) by subexponential
sequences as well, and hence for any sequence, as required by
Definition~\ref{define:basic}. Thus the sporadic capacity is equal
to the capacity.\\
\\
Note that Proposition \ref{propo:hashing1} also clarifies the
questions brought up in Section~\ref{sec:speed}: indeed a
requirement that errors should vanish exponentially fast can be
met for any achievable rate strictly below the capacity. Analogous
results have been presented very recently by M. Hamada
\cite{Ham03}, building on earlier work in
\cite{Ham02b,Ham01,Ham02}.\\
\\
Moreover, it is clear from Proposition \ref{propo:hashing1} that
tolerating finite errors is possible: Since we require the
capacity $Q_\varepsilon(T)$ to be achieved for arbitrarily large
$N$, the second term in Equation (\ref{e:Qforall}) also goes to
zero, and we get the bound
\begin{equation}\label{e:qepsilon}
   Q_\varepsilon(T)\geq Q(T)\geq Q_\varepsilon(T) \, (1-4 \, e \, \varepsilon)\;.
\end{equation}
Hence $\lim_{\varepsilon\to0}Q_\varepsilon(T)=Q(T)$, as claimed.


\ack We would like to thank A. Winter for fruitful discussions,
and A.\,S. Holevo for letting us use his version of the isometric
encoding theorem.\\ Funding from Deutsche Forschungsgemeinschaft
is gratefully acknowledged.


\end{document}